\documentclass[article,12pt]{elsarticle}
\usepackage[a4paper,text={400pt,650pt},centering]{geometry}
\usepackage[utf8]{inputenc}
\usepackage{amsmath}
\usepackage{relsize}
\usepackage{verbatim}
\usepackage{appendix}
\usepackage{mathtools}
\usepackage{tikz}
\usetikzlibrary{decorations.markings}
\usepackage{braket}
\usepackage[new]{old-arrows}
\usepackage{mathtools}

\DeclarePairedDelimiter\floor{\lfloor}{\rfloor}

\usepackage{bbold}
\usepackage{tablefootnote}
\usepackage{float}
\usepackage{amsmath,bm}
\usepackage{enumerate}
\usepackage{pgfplots}
\usepackage{fancyhdr}
\usepgfplotslibrary{fillbetween}
\usetikzlibrary{patterns}
\pgfplotsset{compat=1.12}
\usepackage{mathtools}
\usepackage{graphicx}
\usepackage{caption}
\usepackage{subcaption}

\usepackage{graphicx}
\usepackage[compat=1.0.0]{tikz-feynman}
\usepackage{amsfonts,amssymb,amsthm,epsfig,epstopdf,url,array}
\def\ov{\overline}
\usepackage{braket}

\usepackage{enumerate}
\usepackage{caption}

\usepackage{lineno,hyperref}
\modulolinenumbers[5]

\journal{A}
\biboptions{sort&compress}   %% from Manual

\let\today\relax
\makeatletter
\def\ps@pprintTitle{%
    \let\@oddhead\@empty
    \let\@evenhead\@empty
    \def\@oddfoot{\footnotesize\itshape
         {\sc MPP--2022--288} \hfill\today}%
    \let\@evenfoot\@oddfoot
    }
\makeatother

\def\Tr{\mathop{\rm Tr}}

\theoremstyle{remark}

\newcommand{\pf}{\mathrm{Pf}}

\allowdisplaybreaks

\newcommand{\sect}[1]{ \section{#1} \setcounter{equation}{0} }
\newcommand{\subsect}{\subsection}
\newcommand{\req}[1]{(\ref{#1})}
\def\vev#1{\langle #1 \rangle}

\def\bet{\beta}
\def\fc#1#2{\frac{#1}{#2}}
\def\h{\frac{1}{2}}
\newcommand{\nwc}{\newcommand}
\nwc{\ba}  {\begin{array}}
\nwc{\ea}  {\end{array}}
\nwc{\bdm} {\begin{displaymath}}
\nwc{\edm} {\end{displaymath}}

\nwc{\bea} {\begin{equation}\ba{lcl}}
\nwc{\eea} {\ea\end{equation}}

\nwc{\be} {\begin{equation}}
\nwc{\ee} {\end{equation}}

\nwc{\bda} {\bdm\ba{lcl}}
\nwc{\eda} {\ea\edm}

\nwc{\bc}  {\begin{center}}
\nwc{\ec}  {\end{center}}

\nwc{\ds}  {\displaystyle}

\nwc{\nn} {\nonumber}
\nwc{\nnn} {\nonumber \vspace{.2cm} \\ }
\nwc{\ra}{\rightarrow}
\nwc{\lra}{\longrightarrow}
\def\lf{\left}\def\ri{\right}

\nwc{\p} {\partial}
\def\ap{\alpha'}
\def\lng{\langle}

\def\Mc{{\cal M}}

\def\Rc{{\cal R}}
\def\ov{\overline}

\def\eps{\epsilon}

\def\al{\alpha}
\def\bet{\beta}

\def\IC{{\bf C}}
\def\IP{{\bf P}}
\def\IZ{{\bf Z}}

\def\Ac{{\cal A}}

\def\Sc{{\cal S}}
\def\Rc{{\cal R}}

\def\eps{\epsilon}
\def\al{\alpha}

\def\si{\sigma}
\def\om{\omega}
\def\bet{\beta}\def\bet{\beta}

\begin{document}
\begin{frontmatter}

\title{{\sc  Intersections of Twisted Forms: \\[2mm]
New theories and Double copies}\vskip1.5cm}

\author{Pouria Mazloumi\ and\  Stephan Stieberger}


\address{Max-Planck Institut f\"ur Physik\\Werner--Heisenberg--Institut, 80805 Munich, Germany}

\address{{\it Emails:} pmazlomi@mpp.mpg.de, stieberg@mpp.mpg.de \vskip2cm}

\begin{abstract}

Tree--level scattering amplitudes of particles have a geometrical description in terms of intersection numbers of pairs of twisted differential forms on the moduli space of Riemann spheres with punctures.
We customize a catalog of twisted differential forms containing  both already known and new differential forms.
By pairing elements from this list intersection numbers of various theories can be furnished to compute  their scattering amplitudes. Some of the latter are familiar  through their CHY description, but others are unknown. Likewise, certain pairings give rise to various known and novel double--copy constructions of  spin--two theories. This way we find  double copy constructions for many theories, including higher derivative gravity, (partial massless) bimetric gravity and some more exotic theories.
Furthermore, we present a derivation of  amplitude relations in intersection theory.

\vskip1cm
\end{abstract}


\end{frontmatter}

\newpage

\setcounter{tocdepth}{2}
\tableofcontents
\newpage

\section{Introduction}

The geometry of the moduli space of Riemann spheres with punctures is ubiquitous for a geometrical description of tree--level scattering amplitudes of particles. This geometry appears for
 computing intersection numbers of twisted differential forms, for formulating amplitudes in terms of 
 Cachazo--He--Yuan (CHY) description  and as world--sheet for tree--level closed string amplitudes.
In fact,  the latter  are formulated as a multi--dimensional complex integral over the sphere, which by applying complex analysis can be written in terms of pairs of real integrals subject to some monodromy phases. This procedure gives rise to the  (Kawai--Lewellen--Tye) KLT relations which in turn give rise to an underlying  double copy property intertwining  two open string sectors. This double copy structure can be also be inferred at the level of twisted intersection forms as twisted period relations.
This way by pairing known and new twisted differential forms we may encounter new double copy structures
from twisted intersection numbers.
Intersection numbers provide a geometric description of tree--level $n$--point amplitudes in various 
quantum field theories and likewise a geometric interpretation of the kinematic Jacobi identity \cite{M1}.
In this work we want to further investigate the interrelation between quantum field theory, string scattering amplitudes  and intersection theory. We elaborate both on the computation of the amplitudes and on the construction of possible double copies which arise from those theories.

The computation of scattering amplitudes involving spin--two particles is a tedious challenge in conventional Feynman diagram approach due to the proliferation of diagrams involved. However, since the inception of  Kawai--Lewellen--Tye relations \cite{klt} at the tree--level this task can be 
relegated to the computation of the spin--one sector only by means of doubling the result. In modern language this 
procedure has been formulated as double copy structure of gravity based on color kinematics duality \cite{bcj1,bcj2}. This manipulation replaces the color structure of one gauge theory with the kinematical structure of a second gauge theory. This doubling procedure is not restricted to constructing spin--two interactions
from essentially squaring spin--one interactions but can be applied in principle for any pairs of theories which allow for color kinematics duality. For a detailed account on this see \cite{Bern:2019prr}.
In fact, a peculiar feature of twisted intersection theory is that it exhibits the double copy structure inherited from string theory in a more natural way than QFT amplitudes. In intersection theory constructing double copies
corresponds to paring different twisted intersection forms.

For the construction of differential forms of physical relevance in intersection theory both world--sheet integrands of string amplitudes and the framework of CHY constructions provide fruitful ingredients. Furthermore, an extension of  open superstring interactions through an embedding of the disk onto the sphere allows to accommodate couplings between open and closed strings relevant for twisted forms describing 
EYM amplitudes \cite{MS}. In this work we shall apply this  embedding for the bosonic string theory.

This work is organized as follows. In Section $2$ we introduce some basics of twisted intersection theory. We refer the reader to \cite{MS} for a more detailed introduction to this subject. Here, we primarily   elaborate  on how to extract twisted differentials from CHY integrands and vice versa. In Section 3 we customize a catalog of twisted differential forms containing  both already known and new differential forms. By pairing various differential forms we shall obtain intersection numbers describing the amplitudes of the corresponding theories. For the latter we shall discuss both  
 their Lagrangians and possible CHY representations. In Section 4 by applying the embedding from the disk onto the sphere to  bosonic open string fields  we construct a novel twisted form. The latter allows to construct amplitudes by intersection numbers for a variety of theories including Weyl--YM or $(DF)^2+\phi^3$ theories.
 In Section 5 we elaborate on the generic underlying structure of double copy constructions in intersection theory. We shall present old and new double copies. For the latter  an amplitude description has been unknown previously.
This way we find a double copy construction for many theories, including higher derivative gravity, (partial massless) bimetric gravity and some more exotic theories. In Section \ref{AMPREL} we present a derivation of Kleiss--Kuijf (KK) and Bern--Carrasco--Johansson (BCJ) amplitude relations in intersection theory. 
Finally, the appendices contain some supplementary material.

\sect{CHY integrands for twisted differentials}
\def\Ic{{\cal I}}

Intersection numbers provide a geometric description of tree--level $n$--point amplitudes in various 
quantum field theories.
Selecting a theory requires specifying two twisted differential forms $\varphi_-$ and $\varphi_+$ on the moduli space $\Mc_{0,n}$ of Riemann spheres with $n$ punctures
\begin{equation}\label{Space}
\mathcal{M}_{0,n}=\{(z_1,\ldots,z_i,\ldots,z_n) \in (\IC\IP)^{n-3}\ ,\ i\neq j,k,l \ |\  \mathop{\mathlarger{\forall}}_{m\neq n } z_m\neq z_n \}\ ,
\ee
subject to the three points $z_j,z_k,z_l$ fixed. The space \req{Space} gives rise to the moduli space   of Riemann spheres  with $n$ punctures.  Due to the $SL(2,\IC)$ invariance we consider $(n-3)$--forms $\varphi$ as elements of cohomology equivalence classes
\begin{equation}\label{equivalence}
\varphi\simeq\varphi+\nabla_{\pm\omega} \xi\ ,
\end{equation}
with a rational $n-4$ form $\xi$ and the Gauss--Manin connection $\nabla_{\pm\omega}=d \pm \omega \wedge$ with $d$ the exterior derivative and some closed one--form (twist) $\omega$. In order to make contact to physical amplitudes for the twist  $\omega$ one can select a potential $W$ with $\omega=dW$ as generating function  such that
\begin{equation}\label{potential}
    \omega=\alpha' \sum_{1\leq i, j \leq n } 2p_i p_j\ d\ln(z_i-z_j)\ ,
\end{equation}
with $n$ on--shell momenta $p_i$. The factor $\ap$ is chosen such that $\omega$ is dimensionless. 
Differential forms $\varphi$ with the property (\ref{equivalence}) give rise to classes $[\varphi]$ of twisted forms, cf. Section 2 of Ref. \cite{MS} for further details.
The intersection of the two twisted forms is defined as integral over \req{Space} \cite{MT}
\begin{equation}
    \langle \varphi_+ , \varphi_- \rangle_\omega:=\lf(-\fc{\alpha'}{2\pi i}\ri)^{n-3}\ \int\limits_{\mathcal{M}_{0,n}} \iota_{\omega}(\varphi_+) \wedge \varphi_- \ .
    \label{intersection}
\end{equation}
The map $\iota_{\omega}(\varphi_+)\in H^{n-3}_{+\omega}$ is the restriction of the twisted form over the compact support $H^{n-3}_{\omega,c}(\mathcal{M}_{0,n}, \nabla_{\omega})$  of $H^{n-3}_\omega(\mathcal{M}_{0,n},\nabla_{\omega})$. Otherwise, the integral over the moduli space $\mathcal{M}_{0,n}$ would not be well--defined since the latter is non--compact.
The one--form \req{potential} is related to the scattering equations
\begin{equation}\label{SEQ}
f_k:=\sum_{j\neq k}\frac{p_k p_j}{z_j-z_k}=0\ ,\ \ 1\leq k\leq n\ ,
\end{equation}
which in the limit $\ap\ra\infty$  relate the twisted intersection numbers \req{intersection} to the CHY description.
For this, the intersection form \req{intersection} can be expanded w.r.t. to (twisted) period integrals
\begin{equation}
\langle {C_a\atop\tilde C_b} \otimes KN^{\pm} | \varphi_\pm\rangle=\int\limits_{C_{a},\tilde C_b} KN^\pm\; \varphi_\pm\ \label{Pair}
\end{equation}
relating the twisted cohomology elements $\varphi_\pm\in H_{\pm\omega}^{n-3}$ and their Poincare dual twisted homology cycles $C_a,\tilde C_b\in H^{\pm\omega}_{n-3}(\mathcal{M}_{0,n},KN^\pm)$ 
associated  with the multivalued function $KN^\pm$, respectively.
Then, this expansion allows the saddle point approximation of \req{intersection} for  $\ap\ra\infty$:
\be
\lim_{\alpha' \rightarrow  \infty}\langle \varphi_+,\varphi_- \rangle_\omega=
\alpha'^{\frac{n-3}{2}} 
    \int\limits_{\mathcal{M}_{0,n}} d\mu_n\ \prod_{k=2}^{n-2} \delta(f_k) \  \lim_{\alpha' \rightarrow  \infty} \hat{\varphi}_+\; \hat{\varphi}_-\ .
    \label{alphlim}
\end{equation}

On the other hand, a generic CHY integral describing an $n$--point amplitude of a wide variety of massless QFTs in $d$ the space--time dimensions assumes the following form 
\be\label{Amplitude}
\Ac(n)=\delta^{(d)}\lf(p_1+\ldots p_n\ri)\ \int_{\mathcal{M}_{0,n}} d\mu_n
 \ \widetilde{\Ic}(\{p_i,\tilde\eps_i,z_i\})\ \Ic(\{p_l,\eps_l,z_l\})\ \sideset{}{'}\prod_{l=1}^n\delta(f_l)\ ,
\ee
with the measure
\be
d\mu_n=z_{jk}z_{jl}z_{kl}\prod_{i=1\atop i\notin\{j,k,l\}}^n d^iz\ ,
\ee
which is a degree $n-3$ holomorphic form on $\mathcal{M}_{0,n}$ with $z_j,z_k,z_l$ being three arbitrary marked points fixed by $SL(2,\IC)$ invariance.
Above, the localization of the amplitude at the scattering equation  is achieved through the delta--functions $\delta(f_i)$ with support on \req{SEQ}
There are $(n-3)!$ solutions $\si^{(a)}_k$ of the equations \req{SEQ} and in terms of which the amplitude  \req{Amplitude} becomes
\be\label{Amplitude1}
\Ac(n)=\delta^{(d)}\lf(p_1+\ldots p_n\ri)\ \sum_{a=1}^{(n-3)!}\lf. \Phi^{-1}
 \ \widetilde{\Ic}\ \Ic\ri|_{z_k=\si^{(a)}_k}\ ,
\ee
with $\Phi=\det(\p_j f_a)$ the Jacobian referring to the $n-3$ independent $\delta$--functions.
The numerators $\widetilde{\Ic}$ and $\Ic$ are rational functions of the external data
$\{p_l,\eps_l,z_l\}$ denoting momenta, polarizations $\eps_l$ and points $z_i$ on $\mathcal{M}_{0,n}$.
Specifying a certain theory means  selecting a pair of numerators  $\widetilde{\Ic}$ and $\Ic$. 

We may use the numerators $\widetilde{\Ic}$ and $\Ic$ to define the pair of $n$--forms
\be
\varphi_+=\Ic\ d\mu_n\ \ \ ,\ \ \ \varphi_-=\widetilde{\Ic}\ d\mu_n
\ee
of the twisted cohomology groups $\varphi_+\in H_{+\om}^{n-3}$ and $\varphi_-\in H^{n-3}_{-\om}$, respectively. Then the bilinear pairing through the intersection number \req{intersection} computes the amplitude \req{Amplitude1} as
\be\label{twistCHY}
\Ac(n)=\langle \varphi_+ , \varphi_- \rangle_\omega=
\alpha'^{\frac{n-3}{2}} 
    \int\limits_{\mathcal{M}_{0,n}} d\mu_n\ \prod_{k=2}^{n-2} \delta(f_k) \  \lim_{\alpha' \rightarrow  \infty} \hat{\varphi}_+\; \hat{\varphi}_-\equiv \Ac_{CHY}(n)\ ,
\ee
which in turn by \req{alphlim} is related to the CHY amplitude $\Ac_{CHY}(n)$.
Conversely, we may look for pairs of  twisted forms to derive previously unknown CHY constructions from 
\req{alphlim} like for Weyl--YM and $(DF)^2+\phi^3$.

Color kinematics (CK) duality \cite{bcj1} and subsequently the double copy structure of amplitudes imposes  constraints on the detailed form of the amplitudes. CK duality is realized in many field--theory amplitudes including 
 non--linear--sigma model, (Dirac-)Born--Infeld, Volkov--Akulov and special Galileon theory to be discussed in the next Section. CK duality of gauge theory and the double copy structure of gravity or in general gauge/gravity relations 
 are proven by the  geometry of the corresponding amplitudes  in string theory  by means of monodromy relations on the string world--sheet \cite{klt,stie,Bjerrum-Bohr:2009ulz}. Similar results are  derived at one--loop \cite{Hohenegger:2017kqy,Tourkine:2016bak}, cf. also \cite{Casali:2019ihm,Stieberger:2021daa}.
 Therefore, it is beneficial  to discuss these properties within the framework of intersection theory.
 One of the important features of describing amplitudes  by intersection theory  is that many involved characteristics  of field--theory amplitudes are understood in terms of the  pole structure of their underlying pairs of  twisted forms. In particular,  KLT relations simply follow from linear algebra
within the bilinear combination of twisted forms \cite{Mizera:2017cqs}.
Note, that twisted forms $\varphi$ are assumed to be  elements of cohomology classes, i.e. defined 
subject to \req{equivalence}. As a consequence of the latter equivalence  one can understand and prove the  KK and BCJ relations as cohomology invariant deformations corresponding to the total derivative of $\varphi^{color}$, i.e. as a consequence of \cite{M1}:
\be \label{MIZ}
(d \pm dW \wedge )\,\varphi^{color}_{\pm;n}\simeq 0\ .
\ee
In Subsection \ref{AmpsRel} we shall explicitly demonstrate how to derive KK and BCJ relations from \req{MIZ}. Furthermore, there we shall evidence, that theories with CK duality generically have a common color form in their pairs of twisted forms entering \req{intersection}.  Conversely,  amplitudes of a  theory satisfying BCJ--KK relations imply the existence of $\varphi^{color}$ in the underlying description 
by intersection theory.

In twisted intersection theory the realization  of double copies can be easily achieved if the 
underlying pairs of theories are constructed from pairs of twisted forms  each containing a color factor.
Then by inserting an orthogonal basis into each of the intersection numbers \req{intersection} by means of expanding each twisted form $\varphi_\pm$ w.r.t. an orthonormal basis  of $n$--forms and its dual basis, respectively these two theories are linked by the KLT kernel, cf. Section \ref{NDC} for more details. This procedure has been applied in \cite{MT} to express EYM amplitudes as double copies of YM and gen.YMS, cf. also Section \ref{NDC}.
Equipped with this recipe  as building block   we will find previously unknown  double copies  for  spin--two theories in Section \ref{NDC}.

\sect{Catalog of theories and corresponding twisted forms}

In this section we shall introduce new twisted forms to describe amplitudes of different theories. The latter may have a  field content of spin--zero, spin--one and/or spin--two.
By properly pairing these newly constructed forms with other forms we then build double copies for spin--zero, spin--one or spin--two theories with the latter invariant under linearized diffeomorphisms.
The structure of string scattering amplitudes proves to be indispensable to construct our twisted forms.
The world--sheet monodromy of the punctures of vertex operators in string amplitudes naturally gives rise to
twisted cohomology allowing to specify 
string based twisted forms \cite{M1, MS}. Nevertheless, the proposed twisted forms correspond to
amplitudes of the underlying quantum field theory.
In fact, the CHY formulation of the latter if existent  links the two  descriptions in the $\ap\ra\infty$ limit
\req{twistCHY}.

\subsect{Known theories: their twisted forms and string theory relations}

We start by giving an overview of theories which can be described by twisted intersection forms. In fact, there is a whole catalogue of the latter which can be paired for intersection numbers to  represent a variety  of  different theories   \cite{M1}. 
Many of them also allow for a CHY description. In the following table we list the known theories, their
pair of intersection forms and their CHY representation if there is one.
\begin{table}[H]
\centering
\renewcommand{\thetable}{\arabic{table}a}
\begin{tabular}{|c| c c c c |} 
 \hline
 &&&&\\
  Theory & $\varphi_{-,n}$ & $\varphi_{+,n}$ & CHY representation & Amplitude   \\ [2ex] 
 \hline\hline
 bi-adjoint scalar & $\varphi^{color}_{-,n} $ & $\varphi^{color}_{+,n}$ & $\mathcal{C}_n \mathcal{C}_n  $ & $n$ color scalar \\  [2ex]  
 \hline
 Einstein & $\varphi^{gauge}_{-,n}$  & $\varphi^{gauge}_{+,n}$ & $ \pf' \psi_n \pf' \psi_n$ &  $n$ gravitons \\ [2ex] 
 \hline
 Yang-Mills & $\varphi^{color}_{-,n}$  & $\varphi^{gauge}_{+,n}$ &  $\mathcal{C}_n \pf' \psi_n$ & $n$ gluons \\  [2ex] 
 \hline
 YM+$(DF)^2$ & $\varphi^{color}_{-,n}$  & $\varphi^{bosonic}_{+,n}$ & ?? & \shortstack{$n$ higher derivative \\ gluons} \\   [2ex]  
 \hline
 Einstein--Weyl & $\varphi^{gauge}_{-,n}$  & $\varphi^{bosonic}_{+,n}$ &  ??  & $n$ spin 2 \\    [2ex]  
 \hline
\shortstack{ special Galilean \\ (sGal)} & $\varphi^{scalar}_{-,n}$  & $\varphi^{scalar}_{+,n}$ &  $(\pf' A_n)^4 $ &  \shortstack{ $n$ higher derivative \\ scalars}  \\    [2ex]     
 \hline
 NLSM & $\varphi^{color}_{-,n}$  & $\varphi^{scalar}_{+,n}$  & $\mathcal{C}_n (\pf' A_n)^2$  & $n$ scalars \\    [2ex]  
 \hline
\shortstack{Born--Infeld \\ (BI)} & $\varphi^{scalar}_{-,n}$  & $\varphi^{gauge}_{+,n}$  &  $(\pf' A_n)^2 \, \pf' \psi_n$ & $n$ spin 1 \\   [2ex]  
 \hline
\end{tabular}
\captionof{table}{Known theories, their pairs of twisted forms and their CHY representations.}\label{t1} 
\end{table}
\noindent The twisted forms in Table \ref{t1} assume the following form:
\begin{align}
\varphi^{color}_{\pm,n}&=d\mu_{n}\ \frac{\Tr(T^{c_1}T^{c_2}\ldots T^{c_n})}{(z_1-z_2)(z_2-z_3)\ldots (z_n-z_1)}\equiv \Tr(T^{c_1}T^{c_2}\ldots T^{c_n})\ PT(1,2,\ldots,n)\nn \\
\varphi^{gauge}_{\pm,n}&=d\mu_{n} \int \prod\limits_{i=1}^{n} d \theta_i d\bar{\theta_i}\frac{\theta_k \theta_l}{z_k -z_l} \exp\Bigg\{-\alpha'^2\sum\limits_{i\neq j}\frac{ \theta_i \theta_j p_i \cdot p_j+\bar{\theta}_i \bar{\theta}_j\varepsilon_i \cdot \varepsilon_j +2(\theta_i - \theta_j)\bar{\theta_i}\varepsilon_i \cdot p_j}{z_i-z_j \mp \alpha'^{-1} \theta_i \theta_j}\Bigg\}\ ,\nn\\
\varphi^{scalar}_{\pm,n}&=d \mu_n\; (\pf' A_n)^2=d\mu_{n}\; \frac{ \det {A}_{[kl]}}{(z_k-z_l)^2} \label{listT} \\
\varphi^{bosonic}_{\pm,n}&=d\mu_{n} \lf(\pm\frac{1}{\alpha'}\ri)^{\floor{\frac{n-2}{2}}}\int \prod\limits_{i=1}^{n} d \theta_i d\bar{\theta_i} \exp\Bigg\{\sum\limits^n_{i\neq j}\Bigg(\pm2  \alpha'\frac{\theta_i \bar{\theta}_i  \varepsilon_i \cdot p_j }{z_i-z_j}+\frac{\theta_i \bar{\theta}_i \theta_j \bar{\theta}_j \varepsilon_i \cdot \varepsilon_j}{(z_i-z_j)^2}\Bigg)\Bigg\} \nn\\
&\stackrel{\alpha' \rightarrow \infty}{\longrightarrow}\  W_{\underbrace{11...1}_{n}}\ d\mu_{n}\ . \nn
\end{align}
The matrices in the CHY representation are defined in the same way as \cite{chy}:
\begin{equation}
\psi_n= \begin{pmatrix} A & -C^T \\ C & B \end{pmatrix}\ ,\label{psi}
\end{equation} 
with the three matrices:
\begin{align}  A_n= \begin{cases}  0\ , \,\,\,\,\,\,\,\, i=j \\
                    \frac{p_i p_j}{z_i-z_j}\ , \,\,\,\,\,\,\,\, i \neq j  
       \end{cases},\  \,\,
& C_n = \begin{cases}  -\sum\limits_{k \neq i}^{n}\frac{\varepsilon_i p_k}{z_i-z_k}\ , \,\,\,\,\,\,\,\, i=j \\
                   \frac{\varepsilon_i p_j}{z_i-z_j}\ ,  \,\,\,\,\,\,\,\,i \neq j 
       \end{cases}, \,\,
 & B_n= \begin{cases}  0\ , \,\,\,\,\,\,\,\, i=j\ , \\
                    \frac{\varepsilon_i  \varepsilon_j}{z_i-z_j}\ ,\,\,\,\,\,\,\,\, i \neq j\ .
       \end{cases} \label{psi1}
 \end{align}
Furthermore, in the definition of $\varphi^{scalar}_{\pm;n}$ there appears  the reduced Pfaffian for the
matrix~$A_n$
\begin{equation}\label{redPF}
\pf'A_n \equiv \frac{(-1)^{k+l}}{z_k-z_l}\; \pf A_{[kl]}\ ,
\end{equation}
with the index $k,l$ denoting removals of  rows $k,l$ and columns $k,l$.
Furthermore, by using $\pf M^2=\det M$ we have the following relation: 
\begin{equation}
\frac{\det A_{[kl]}}{z^2_{kl}}=  (\pf' A_n)^2\ . 
\end{equation}

In the following we review some basics about the special Galilean, non--linear sigma model and Born--Infeld theories displayed in Table \ref{t1}.
The Einstein--Weyl theory will be discussed in Section \ref{NDC}.

\begin{itemize} 

\item Special Galilean (sGal)

Galilean theories in $d$ dimension are defined as scalar effective field theories involving higher derivative interactions in the potential \cite{galrev,ELVANG2} namely
\begin{equation}
\begin{aligned}
 & \mathcal{L}_{ Gal}= \frac{1}{2}\; \partial^\mu \phi \partial_\mu \phi+\sum\limits_{n=3}^{d-1} g_n \,(\partial \phi)^2 (\partial_\mu \partial^\mu \phi)^{n-2}\ , \label{sgal}
\end{aligned}
\end{equation}
or variants thereof \cite{Cachazo2014}.
The Galilean theory is called special if it features a $\IZ_2$ symmetry. The amplitude of $n$ scalars in this theory  \cite{Cachazo2014} can be written in terms of the intersection number \req{intersection} leading to the correct CHY formulation  \req{twistCHY} in the following way:
 \begin{equation}
\Ac_{sGal}(n) =\lim_{\ap\ra\infty}\langle\varphi_{-,n}^{scalar},\varphi_{+,n}^{scalar}\rangle_\omega =\int\limits_{\mathcal{M}_{0,n}} d\mu_n\ \prod_{k=2}^{n-2} \delta(f_k) \,\, (\pf' A_n)^4\ .
     \label{CHY gal}
     \end{equation}

\item Non--linear sigma model (NLSM)

The non--linear sigma model (NLSM) is defined as a theory with scalars $\Phi$ together with an embedding onto a manifold $M$ with a non--linear interaction potential $V(\Phi)$. The theory for $N$ scalars  fields   $\Phi$ satisfying an $U(N)$ symmetry can be written in terms of  Cayley
parametrization \cite{Cachazo2014}:
\begin{equation}
\mathcal{L}_{NLSM}=\frac{1}{8 \lambda^2} \Tr \lf\{\;\partial_\mu U(\Phi) \partial^\mu U(\Phi)\; \ri\}\ , \ee
with:  
\be
U(\Phi):=(\bm{1}+\lambda \Phi)\;(\bm{1}-\lambda \Phi)^{-1}\ .\label{Cayley}
\end{equation}
The field $\Phi$ may be written in the adjoint representation  as $\Phi=\phi^I T^I$ with $T^I$ the generators of $U(N)$. Upon expanding $U(\Phi)$ in terms of $\Phi$ yields the usual scalar kinetic term. 
 The scattering amplitude of this theory involving $n$ scalars  \cite{Cachazo2014} can be given in terms of the intersection number \req{intersection} leading to the correct CHY formulation \req{twistCHY} as:
 \be
\Ac_{NLSM}(n)  =\lim_{\ap\ra\infty}\langle\varphi_{-,n}^{color},\varphi_{+,n}^{scalar}\rangle_\omega=\int\limits_{\mathcal{M}_{0,n}} d\mu_{n}\ \prod_{k=2}^{n-2} \delta(f_k) \,\, 
\mathcal{C}_n\, (\pf' A_n)^2\ .   \label{CHY NLSM}
\ee

\item Born--Infeld theory (BI)

The Born--Infeld theory is the non--linear generalization of Maxwell theory \cite{Tse99}. The Lagrangian for this theory in $d=4$ is given by the non--linear interaction
\begin{equation}
\begin{aligned}
 \mathcal{L}_{BI}= \ell^{-2}\Bigg(\sqrt{-{\det}_{4}(\eta_{\mu\nu}+\ell F_{\mu\nu}) }-1\Bigg)\ ,
\end{aligned}\label{BIA}
\end{equation}
where the scale $\ell$ can be related to the inverse string tension $\ap$ as $\ell=2 \pi \alpha'$. The $n$--point amplitude of this theory \cite{Cachazo2014} can be expressed in terms of the intersection number \req{intersection} leading to the CHY representation of this theory \req{twistCHY}  in terms of the following localization integral: 
\begin{equation}
\Ac_{BI}(n)  =\lim_{\ap\ra\infty}\langle\varphi_{-,n}^{scalar},\varphi_{+,n}^{gauge}\rangle_\omega=\int\limits_{\mathcal{M}_{0,n}} d\mu_{n}\ \prod_{k=2}^{n-2} \delta(f_k) \,\, \pf' \psi_n\; (\pf' A_n)^2\ .
       \label{CHY BI}
     \end{equation}

\end{itemize}

Actually, all twisted forms in Table \ref{t1} are derived from string theory, which in turn  also motivated  the corresponding CHY descriptions. In our previous work  \cite{MS} we were mainly concerned with the two twisted forms $\varphi^{color}$ and $\varphi^{gauge}$. The latter refers to superstring theory and is relevant to describe both SYM and GR amplitudes. Therefore, it is natural to 
discuss the twisted form $\varphi^{bosonic}$ stemming from bosonic string theory. While the 
partial gluon subamplitudes $A_{YM}(1,\ldots,n)$ are  the building blocks for the open superstring amplitudes \cite{Mafra:2011nv,Mafra:2011nw,Broedel:2013tta}, i.e.
\be
\Ac_{open\atop superstring}(1,\pi(2,\ldots,n-2),n-1,n)=\sum_{\rho,\tau\in S_{n-3}}Z(\pi|\tau)\ S[\rho|\tau]_1\ 
A_{YM}(1,\rho(2,\ldots,n-2),n-1,n) \ ,\label{OSS}
\ee
with the iterated disk integrals $Z(\pi|\tau)=Z(1,\pi(2,\ldots,n-2),n-1,n|1,\tau(2,\ldots,n-2),n,n-1)$, 
the same role is played by the subamplitudes $B(1,\ldots,n)$ for $(DF)^2+YM$ gauge theory for the open string \cite{Aze1}, i.e.:
\be
\Ac_{open\atop bosonic\ string}(1,\pi(2,\ldots,n-2),n-1,n)=\sum_{\rho,\tau\in S_{n-3}}Z(\pi|\tau)\ S[\rho|\tau]_1\ 
B(1,\rho(2,\ldots,n-2),n-1,n) \ .\label{OBS}
\ee 
Above the KLT kernel $S[\rho|\tau]_1$ is given by  the $k!\times k!$--matrix \cite{Bern:1998sv,Bjerrum-Bohr:2010pnr}
\be
S[\sigma|\rho]_\ell:=S^{(0)}[\, \si(1,\ldots,k) \, | \, \rho(1,\ldots,k) \, ]_1=\prod_{t=1}^{k} \lf(p_1 p_{t_\si}+\sum_{r<t}p_{r_\si}p_ {t_\si} \theta(r_\si,t_\si)\ri)\ ,\label{Kernel}
\ee
with $j_\si=\si(j)$ and  $\theta(r_\si,t_\si)=1$
if the ordering of the legs $r_\si,t_\si$ is the same in both orderings
$\si(1,\ldots,k)$ and $\rho(1,\ldots,k)$, and zero otherwise.
By comparing the two expressions \req{OSS} and \req{OBS}  it is natural to construct from the latter the  twisted form $\varphi^{bosonic}$  relevant to $(DF)^2+YM$ theory. The latter involves couplings of YM and $(DF)^2$ theory.
The bosonic field theory of the latter is defined by the following Lagrangian \cite{Joh1} 
\begin{equation}
\begin{aligned}
 \mathcal{L}_{YM+(DF)^2} &=\frac{1}{2}\;(D_{\mu}F^{a\,\mu\nu})^2-\frac{g}{3}\; F^3+\frac{1}{2}\ (D_{\mu} \varphi^a)^2+\frac{g}{2}\;  C^{\alpha \, ab} \varphi^{\alpha} F^{a}_{\mu\nu}F^{b \,\mu\nu} \\
 & + \frac{g}{3!}\; d^{\alpha \beta \gamma} \varphi^{\alpha} \varphi^{\beta} \varphi^{\gamma} -\frac{1}{2}\; m^2 (\varphi^\alpha)^2 -\frac{1}{4}\;m^2 F^2\ ,
\end{aligned}\label{LDFYM}
\end{equation}
with the field content comprising a massless gluon, a massive gluon and a massive scalar.
The Lagrangian of $(DF)^2$--theory only will be displayed in Eq. (\ref{LDF}). To the latter  the YM part is added as a mass deformation, which in total gives rise to the Lagrangian \req{LDFYM}.
As a consequence  there are mass terms proportional to $m^2$ for the fields $\varphi^\alpha$ and $F_{\mu\nu}^a$. Within bosonic string theory
this mass is related to the string tension as
\be\label{tachyon}
m^2=-\ap^{-1}\ ,
\ee
accounting for the string tachyonic modes representing both the massive gluon and the massive scalar.
In fact, for $\ap\ra0$ we recover pure YM theory (after multiplying the Lagrangian by an overall factor of $m^{-2}=\ap$), while for $\ap\ra\infty$ we obtain $(DF)^2$ theory.
Hence, after taking the limit  $\alpha' \rightarrow \infty$ the mass \req{tachyon} goes to zero and the  Lagrangian boils down to  $(DF)^2$--theory, i.e. 
\begin{equation}
\lim_{\alpha' \rightarrow \infty} \mathcal{L}_{YM+(DF)^2}(\alpha')\simeq\mathcal{L}_{DF^2}\ , \label{masslim}
\end{equation}
with the latter referring to (\ref{LDF}).
This is consistent with the CHY construction, which works only for massless theories. Likewise, the $\ap\ra\infty$ limit of the intersection number \req{intersection} discards the mass term and reproduces the  massless $(DF)^2$ theory.

In \req{listT} for $\varphi^{bosonic}$ we have introduced the function $W$, which arises  in its $\alpha' \rightarrow\infty$ limit. More concretely,  the latter is used e.g. in \cite{He:2016iqi} and the set $L$ of indices of the function $W_{L}$ refer to a product of Lam--Yao cycles:
\begin{equation}
 W_{\underbrace{11...1}_{n}}=\prod_{i=1}^n C_{ii}= \prod\limits_{i=1}^n \Bigg( \sum\limits_{j=1 \atop j \neq i}^n \frac{\varepsilon_i \cdot p_j}{z_{ij}}\Bigg)\ .
 \label{listT2}
\end{equation}
Notice,  that $W_{11...1}$ is produced from  the twisted form $\varphi^{bosonic}$ in the limit $\alpha'\rightarrow\infty$. In this limit the twisted form  $\varphi^{bosonic}_{\pm,n}$ can be used for describing 
$(DF)^2$ theory. The limit $\alpha' \rightarrow\infty$ removes in $\varphi^{bosonic}_{\pm,n}$  all the  contractions $\eps_i\eps_j$ in agreement with the absence of those terms in $(DF)^2$ theory.
Therefore, we define the following form: 
\begin{equation}
\varphi^{Bosonic}_{\pm,n}=d\mu_{n}\; \lf(\pm\frac{1}{\alpha'}\ri)^{\floor{\frac{n-2}{2}}} \int \prod\limits_{i=1}^{n} d \theta_i d\bar{\theta_i}\; \exp\Bigg\{\sum\limits^n_{i\neq j}\Bigg(\pm 2 \alpha'\frac{\theta_i \bar{\theta}_i  \varepsilon_i \cdot p_j  }{z_i-z_j}\Bigg)\Bigg\}\ = W_{\underbrace{11...1}_{n}}\ d\mu_{n} \ . \label{phiB}
\end{equation} 
The last equality is up to an overall sign. On the other hand, for the CHY representation of $YM+(DF)^2$ theory we cannot use the limit $\alpha' \rightarrow\infty$  since this theory is massive with the mass given by \req{tachyon}. As a consequence  taking the limit $\alpha' \rightarrow\infty$ removes the mass term in \req{LDFYM} referring to the YM part of this theory. As discussed before, 
 in this limit only the   CHY representation of $(DF)^2$--theory is reproduced.  

A similar situation applies to Einstein--Weyl theory, cf. Subsection \ref{WEW}. In this case the mass term refers to the Einstein term and one obtains only the conformal gravity (i.e. Weyl) amplitude in the  limit $\alpha' \rightarrow\infty$.  Additionally, this also relates to the notion of massive CHY amplitudes. To conclude, while we do have representations in twisted intersection theory for amplitudes of both $YM+(DF)^2$ and $(DF)^2$ theories we only have CHY representations for the latter.
The same is true for conformal gravity and Einstein--Weyl theories, cf. Section \ref{NDC}.
 
In this work one of our goals  is to both complete and extend the Table \ref{t1}. Furthermore, we want to understand how these constructions relate
 to string theory, color--kinematics duality and double copies.

 \subsection{More theories from  the Einstein Yang--Mills form $\tilde{\varphi}^{EYM}_{\pm,n;r}$}

  The first extension of Table \ref{t1} has been accomplished in our publication \cite{MS}. There,  we
 have shown, that one can extend the twisted intersection description \req{intersection} to Einstein Yang--Mills (EYM) amplitudes by introducing an embedding formalism. Concretely,
 we introduced the twisted form:
\begin{equation}
\begin{aligned}
\widetilde{\varphi}^{EYM}_{\pm,n;r}=& d\mu_{n+r}\,\,\mathcal{C}(1,2,3,...,n)\int  \prod\limits_{ i\in\Sc_r}   d \theta_i d\bar{\theta_i} \ \exp\Bigg\{\frac{1}{2} \ap^2\sum\limits_{i,j\in \Sc_r} \begin{pmatrix} \theta_i \\ \bar{\theta}_i  \end{pmatrix}^t \Psi_r  \,\, \begin{pmatrix} \theta_j \\ \bar{\theta}_j  \end{pmatrix}\Bigg\}\Bigg|_{\ov{\zeta_l} \rightarrow \zeta_l}\\
   &  \times \exp\Bigg\{\pm \frac{1}{2} \ap \sum\limits_{i,j\in \Sc_r} \frac{\theta_i \theta_j \ov{\theta}_i \ov{\theta}_j (\xi_i \cdot \xi_j)}{(\ov{\zeta}_i-\ov{\zeta}_j)^2}  \Bigg\}\ .
\end{aligned} \label{TWF1}
\end{equation}
With \req{TWF1} we have the following table
\begin{table}[H]
\addtocounter{table}{-1}
  \renewcommand{\thetable}{\arabic{table}b}
  \centering
\begin{tabular}{|c| c c c c|} 
 \hline
 &&&&\\
  Theory & $\varphi_-$ & $\varphi_+$ & CHY representation &  Amplitude   \\ [2ex] 
 \hline\hline
EYM & $\tilde{\varphi}^{EYM}_{-,n;r}$ & $\varphi^{gauge}_{+,n+r}$ & $\mathcal{C}_n\; \pf \Psi_{S_r}\;  \pf' \psi_{n+r} $ & $r$ gravitons, $n$ gluons \\  [2ex]  
 \hline
\end{tabular}
\captionof{table}{Twisted forms for EYM and its CHY representation.}\label{t1b}
\end{table}
\noindent
 to construct EYM amplitudes involving $n$ gluons and $r$ gravitons as:
\begin{equation}
\begin{aligned}
  \Ac_{EYM}(n;r) &=\lim_{\alpha' \rightarrow  \infty}\langle \tilde{\varphi}^{EYM}_{-,n;r},{\varphi}^{gauge}_{+,n+r} \rangle_\omega=
    \int\limits_{\mathcal{M}_{0,n+r}} d\mu_{n+r}\ \prod_{k=2}^{n+r-2}\ \delta(f_k) \  \lim_{\alpha' \rightarrow  \infty} \hat{\tilde{\varphi}}^{EYM}_{-,n;r} \,\, \hat{\varphi}^{gauge}_{+,n+r} \\
    &  =\int\limits_{\mathcal{M}_{0,n+r}} d\mu_{n+r}\ \prod_{k=2}^{n+r-2} \delta(f_k) \   \mathcal{C}_n\,  \pf \Psi_{S_r} \, \pf' \psi_{n+r}\ .
     \end{aligned}    \label{Amp EYM}
     \end{equation}
 Actually, we can go further by looking at other theories, which have similar CHY structures than EYM. The first extension originates
 from paring  $\tilde{\varphi}^{EYM}_{-,n;r}$ with other twisted forms from \req{listT} and comparing \req{alphlim} with the corresponding CHY representation \cite{Cachazo2014}. This way it is straightforward to construct amplitudes for generalized Yang--Mills Scalar (gen.YMS) and extended Dirac Born--Infeld (ext.DBI) supplementing our Table \ref{t1} by the following content:

\begin{table}[H]
\hskip-0.75cm%\centering
\begin{tabular}{|c| c c c c|} 
 \hline
 &&&&\\
  Theory & $\varphi_-$ & $\varphi_+$ & CHY representation &  Amplitude    \\ [2ex] 
 \hline\hline
\shortstack{Generalized Yang-Mills scalar \\ (gen.YMS)} & $\tilde{\varphi}^{EYM}_{-,n;r}$ & $\varphi^{color}_{+,n+r}$ & $\mathcal{C}_n \pf \Psi_{S_r}  \mathcal{C}_{n+r}$ & \shortstack{ $r$ gluons \\ $n$ color scalars }  \\  [2ex]  
 \hline
\shortstack{Extended Dirac Born-Infeld \\ (ext.DBI)} & $\tilde{\varphi}^{EYM}_{-,n;r}$ & $\varphi^{scalar}_{+,n+r}$ & $\mathcal{C}_n \pf \Psi_{S_r}  (\pf' A_{n+r})^2 $  & \shortstack{ $r$ gluons \\ $n$ higher derivative scalars }  \\  [2ex]  
 \hline
\end{tabular}
\captionof{table}{Additional theories that can be described through new twisted form $\widetilde{\varphi}^{EYM}$.}\label{t2} 
\end{table}
\noindent In order to verify  our Table \ref{t2} we compute the intersection numbers \req{intersection} in the limit $\ap\ra\infty$ given by  (\ref{alphlim}) involving the  pairs of  twisted forms
for gen.YMS and ext.DBI  and compare the results with their corresponding CHY representations.

In the following, we shall give some details about the two theories, which have appeared in the previous Table \ref{t2}.

\begin{itemize}
\item Generalized Yang--Mills scalar (gen.YMS)

Generalized Yang--Mills Scalar (gen.YMS) is described by YM gauge theory  coupled to scalars $\phi^{a\tilde a}$ with both one color $a$ and one flavour index $\tilde a$ \cite{Cachazo2014} i.e.:
\begin{equation}
\begin{aligned}
& \mathcal{L}_{gen.YMS}=-\frac{1}{4}F_a^{\mu\nu} F^{a\mu\nu}-\frac{1}{2} (D_\mu \phi^{a\tilde a})^2+\lambda\ f^{abc} f^{\tilde a\tilde b\tilde c}\ \phi^{a\tilde a} \phi^{b\tilde b} \phi^{c\tilde c}  \ .  \label{LgenYMS} 
\end{aligned}
\end{equation}
 Actually, if the second set of labels $\tilde a$ also represents color indices we have two color groups
 and obtain  pure YM plus cubic biadjoint scalar theory $YM+\phi^3$.
 Furthermore, without the three--point interaction $ \lambda\ra0$ in the Lagrangian \req{LgenYMS} we are obtaining Yang Mills Scalar (YMS) theory.
As in the definition of \req{LgenYMS} the group structure of the theory at hand is extended by a flavour group for multiple scalars $\phi^{a\tilde a}$. However,  here we shall consider
only one scalar in the adjoint representation of the gauge group with a single trace for the flavour group.
As proposed above, with (\ref{alphlim}) we can construct the CHY representation of the gen.YMS amplitude involving $r$ gluons and $n$ scalars:
\begin{equation}
\begin{aligned}
  \Ac_{gen.YMS}(n;r) &=\lim_{\alpha' \rightarrow  \infty}\langle \varphi^{color}_{-,n+r},\tilde{\varphi}^{EYM}_{+,n;r} \rangle_\omega=
    \int\limits_{\mathcal{M}_{0,n+r}} d\mu_{n+r}\ \prod_{k=2}^{n+r-2} \delta(f_k) \  \lim_{\alpha' \rightarrow  \infty} \hat\varphi^{color}_{-,n+r} \,\, \hat{\tilde{\varphi}}^{EYM}_{+,n;r}\\
    &  =\int\limits_{\mathcal{M}_{0,n+r}} d\mu_{n+r}\ \prod_{k=2}^{n+r-2} \delta(f_k) \   \mathcal{C}_n \ \pf \Psi_{S_r}\  \mathcal{C}_{n+r}\ .
     \end{aligned}    \label{Amp GYMS}
     \end{equation}

\item Extended Dirac Born--Infeld

There are two extensions of the the Born--Infeld (BI) theory, which we will discuss throughout out this work. Firstly, there is the Dirac Born--Infeld (DBI) theory. The latter is defined as an scalar extension of the BI theory \req{BIA} with $n$ scalars and the Lagrangian is given by \cite{Cachazo2014}:  
\begin{equation}
\begin{aligned}
 \mathcal{L}_{DBI}= \ell^{-2}\ \Bigg(\sqrt{-{\det}_{4}(\eta_{\mu\nu}-\ell^2 \partial_{\mu} \phi^{I} \partial_{\nu} \phi^{I}+\ell F_{\mu\nu}) }-1\Bigg)\ .
\end{aligned}\label{DBIaction}
\end{equation}
Furthermore, one can  extend this theory by  generalizing the scalar kinetic term by the  function $U(\Phi)$ given in \req{Cayley} and leading to extended DBI theory (ext.DBI) with the corresponding  Lagrangian
\begin{equation}
\begin{aligned}
 \mathcal{L}_{ext.DBI}= \ell^{-2}\ \Bigg(\sqrt{-{\det}_{4}(\eta_{\mu\nu}-\frac{\ell^2}{4\lambda^2} \Tr (\partial_{\mu} U^{\dagger} \partial_{\nu} U)-\ell^2W_{\mu\nu}+ \ell F_{\mu\nu})}-1\Bigg)\ ,
\end{aligned}\label{LagextDBI}
\end{equation}
and: 
\begin{equation}
\begin{aligned}
W_{\mu\nu}& =\sum\limits_{m=1}^{\infty} \sum\limits_{k=0}^{m-1} \frac{2(m-k)}{2m-1}\;\lambda^{2m+1}\; \Tr (\partial_{[\mu} \Phi \,\, \Phi^{2k} \partial_{\nu]}\Phi \,\,\Phi^{2(m-k)-1})\ .
\end{aligned}
\end{equation}
For $\lambda\ra0$ the expression \req{LagextDBI} yields the   DBI action \req{DBIaction}, while we recover NSLM for $\ell\ra0$, respectively.
With  (\ref{alphlim})  the amplitude for ext.DBI involving $r$ gluons and $n$ scalars can be constructed as
\begin{equation}
\begin{aligned}
  \Ac_{ext.DBI}(n;r) &=\lim_{\alpha' \rightarrow  \infty}\langle \varphi^{scalar}_{-,n+r},\tilde{\varphi}^{EYM}_{+,n;r} \rangle_\omega=
    \int\limits_{\mathcal{M}_{0,n+r}} d\mu_{n+r}\ \prod_{k=2}^{n+r-2} \delta(f_k) \  \lim_{\alpha' \rightarrow  \infty} \hat\varphi^{scalar}_{-,n+r} \,\, \hat{\tilde{\varphi}}^{EYM}_{+,n;r}\ ,\\
    &  =\int\limits_{\mathcal{M}_{0,n+r}} d\mu_{n+r} \ \prod_{k=2}^{n+r-2} \delta(f_k) \;   \mathcal{C}_n  \;\pf \Psi_{S_r} \; (\pf' A_{n+r})^2\ ,
     \end{aligned}    \label{Amp D BI}
     \end{equation}
     which is the CHY amplitude given in \cite{Cachazo2014}. The double copy structure
     of BI, DBI and ext.DBI theories will be discussed in Section \ref{NDC} resulting in the representations \req{DC7}, \req{DC3} and  (\ref{DC5}), respectively.

\end{itemize}

\subsection{Theories involving the Einstein--Maxwell form  $\tilde{\varphi}^{EM}_{\pm,n;r}$}

Above we have used the twisted form \req{TWF1} of EYM to compute the intersection number  (\ref{alphlim}) for the amplitudes of   EYM, gen.YMS and ext.DBI theories given in \req{Amp EYM}, \req{Amp GYMS} and \req{Amp D BI}, respectively.  While for the latter a known CHY representation exists \cite{Cachazo2014},  their  corresponding twisted forms had not been constructed so far.
Similarly, the twisted form for Einstein--Maxwell (EM), i.e. Einstein gravity with  an $U(1)^m$ gauge group, as it arises from compactification of $m$ dimensions, can be used to construct
other amplitudes for  theories which interact with Maxwell theory.    One would expect these theories to be defined in some limit of the Yang--Mills theory. We organize them in the following Table \ref{t3} based on \cite{Cachazo2014}.
\begin{table}[H]
\centering
\begin{tabular}{|c|  c c |} 
 \hline
&&\\
  Theory &  CHY representation &  Amplitude   \\ [2ex] 
 \hline\hline
 \shortstack{Einstein--Maxwell \\ (EM) } &  $ \pf X_n\, \pf' \Psi_{S_{r:n}}\; \pf' \psi_{n+r} $ & \shortstack{$r$ graviton \\ $n$ photon }  \\  [2ex]  
 \hline
 \shortstack{Dirac Born--Infeld \\ (DBI)} & $  \pf X_n\, \pf' \Psi_{S_{r:n}}\;(\pf' A_{n+r})^2 $  & \shortstack{$r$ gluons \\ $n$ color scalars }    \\   [2ex] 
 \hline
\shortstack{Yang-Mills scalar \\ (YMS)} &   $ \pf X_n\, \pf' \Psi_{S_{r:n}}\;\mathcal{C}_{n+r} $ & \shortstack{$r$ gluons \\ $n$ color scalars }   \\  [2ex] 
 \hline
\end{tabular}
\captionof{table}{Known theories yet without twisted from description.} \label{t3}
\end{table}
\noindent The matrix  $\Psi_{S_{r:n}}$ is similar to the object $\Psi_{S_r}$ introduced in \cite{MS}. 
The latter is a $(2r)\times(2r)$--matrix with only those indices  included, which refer to the  highest spin particles of  the theory under consideration.  On the other hand, the object $\Psi_{S_{r:n}}$ is a 
$(2r+n)\times (2r+n)$--matrix with an additional (half) sector contributing for the lower spin particles.
More concretely, for EM we have  \cite{Cachazo2014}
\be
\Psi_{S_{r:n}}=\begin{pmatrix}
A_{ab}&A_{aj}&(-C^t)_{ab}\\
A_{ib}&A_{ij}&(-C^t)_{aj}\\
C_{ab}&C_{aj}&B_{ab}
\end{pmatrix}
\ee
described by the sets of graviton indices $a,b\in\{1,\ldots,r\}$ and the photon (gluon) indices $i,j\in\{1,\ldots,n\}$. In cases of DBI and YMS  we have the sets of gluon indices $a,b\in\{1,\ldots,r\}$ and the scalar indices $i,j\in\{1,\ldots,n\}$.

The only common part in the CHY representation of all three theories is comprised by the Pfaffian   $\pf X_n$ of the $n \times n$ matrix $X$ with matrix elements:
\begin{equation}
\begin{aligned}
X_{ab}=\begin{cases} &  \frac{1}{z_a-z_b} \hspace{.1cm} \quad a \neq b\ ,  \\
& 0 \qquad\quad  a=b\ .
\end{cases}
\end{aligned} 
\end{equation}
This Pfaffian describes a correlator involving an even number of $n$ fermions, i.e.:
\be\label{fermionicPf}
\vev{\psi(z_1)\ldots\psi(z_n)}=\pf X_n\ .
\ee
Therefore, the number $n$ of photons must be even. The twisted form of EM can be represented by the following $(n+r)$--form:
\begin{equation}
\begin{aligned}
\tilde{\varphi}^{EM}_{\pm,n;r}&  = d\mu_{n+r} \,\, \int  \prod\limits_{ i\in\Sc_r}   d \theta_i d\bar{\theta_i} \, \frac{\theta_k \theta_l}{z_k -z_l}\, \exp\Bigg\{\frac{1}{2} \ap^2\sum\limits_{i,j\in \{r,n\}} \begin{pmatrix} \theta_i \\ \bar{\theta}_i  \end{pmatrix}^t \Psi_{S_{r:n}}  \,\, \begin{pmatrix} \theta_j \\ \bar{\theta}_j  \end{pmatrix}\Bigg\}\;  \times \pf X_{n}\ ,
\end{aligned} \label{EM}
\end{equation}
where we have used the Pfaffian in equation (\ref{fermionicPf}). Together with the twisted form for YM $\varphi^{gauge}_{-,n+r}$ we can construct the Einstein-Maxwell amplitude 
\begin{equation}
\begin{aligned}
  \Ac_{EM}(n;r) &=\lim_{\alpha' \rightarrow  \infty}\langle \varphi^{gauge}_{-,n+r},\tilde{\varphi}^{EM}_{+,n;r} \rangle_\omega=
    \int\limits_{\mathcal{M}_{0,n+r}} d\mu_{n+r}\ \prod_{k=2}^{n+r-2}\ \delta(f_k) \  \lim_{\alpha' \rightarrow  \infty} \hat{\varphi}^{gauge}_{-,n+r} \,\, \hat{\tilde{\varphi}}^{EM}_{+,n;r}\ ,\\
    &  =\int\limits_{\mathcal{M}_{0,n+r}} d\mu_{n+r}\ \prod_{k=2}^{n+r-2}\ \delta(f_k) \  \pf X_{n} \,\, \pf' \Psi_{S_{r:n}}\;\,\, \pf'\psi_{n+r}\ ,\\
     \end{aligned}    \label{Amp EM}
     \end{equation}
     involving $r$ gravitons and $n$ photons.
Similar constructions can be established for the two other theories displayed in Table \ref{t3}. In total, we have the following pairings of twisted forms:
\begin{table}[H]
\centering
\begin{tabular}{|c| c c c |} 
 \hline
&&&\\
  Theory & $\varphi_-$ & $\varphi_+$ & CHY representation  \\ [2ex] 
 \hline\hline
 EM & $\varphi^{gauge}_{-,n+r}$ & $\tilde{\varphi}^{EM}_{+,n;r}$ & $ \pf X_n\; \pf' \Psi_{S_{r:n}}\; \pf' \psi_{n+r} $ \\  [2ex]  
 \hline
 DBI & $\varphi^{scalar}_{-,n+r}$  & $\tilde{\varphi}^{EM}_{+,n;r}$ & $  \pf X_n\;  \pf' \Psi_{S_{r:n}}\; (\pf' A_{n+r})^2 $   \\   [2ex] 
 \hline
 YMS & $\varphi^{color}_{-,n+r}$  & $\tilde{\varphi}^{EM}_{+,n;r}$ &  $ \pf X_n\;  \pf' \Psi_{S_{r:n}}\; \mathcal{C}_{n+r} $  \\  [2ex] 
 \hline
\end{tabular}
\captionof{table}{Theories which can be described by  $\tilde{\varphi}^{EM}_{\pm,n;r}$.}\label{t4}
\end{table}

\sect{New and old theories from a novel  twisted form}

So far we have only considered those extensions of  Tables \ref{t1} and  \ref{t1b}, which refer to known theories. This has been achieved by paring  known twisted forms with either the EYM form  $\tilde{\varphi}^{EYM}_{\pm,n;r}$  or the EM form $\tilde{\varphi}^{EM}_{\pm,n;r}$ given in \req{TWF1} and \req{EM}, respectively.  However, we want to extend and generalize our considerations by constructing  new twisted forms relevant for   different types of spin--two theories, e.g. Weyl, conformal or  $R^3$ gravity. These theories allow for CHY representations related to constructions in string theory \cite{Aze1} or ambitwistor string theory \cite{Aze}. The theories and their CHY representations we shall be concerned with are displayed in Table \ref{tt4}.

\subsect{Novel twisted form  from  disk embedding}
\def\Nc{{\cal N}}

The  function
\begin{equation}
\widetilde{W}_{\underbrace{11\ldots1}_{r}}=\prod\limits_{i \in \Rc} \Bigg(\sum\limits_{j \in \Nc}\frac{\varepsilon_i \cdot p_j}{z_{ij}} + \sum\limits_{j \in \Rc \atop j \neq i} \frac{\varepsilon_i \cdot p_j}{z_{ij}}\Bigg)\ ,\label{Weq}
\end{equation} 
which has been introduced in \citep{Aze1},  generalizes \req{listT2} and describes the Weyl--YM theory.
The set $\Nc$ encompasses all gluon labels while the set $\Rc$ comprises all graviton labels. In the  object \req{Weq} the diagonal elements of the $C$--block \req{psi1} entering the matrix $\psi_n$ in \req{psi} appear. 
A special limit of the function \req{Weq} is exhibited in (\ref{listT}), which gives rise to \req{listT2} and stems from the  twisted form $\varphi^{bosonic}_{\pm,n}$ originating from the open bosonic string theory.
Here we want to construct the new  twisted form $\widetilde{\varphi}^{Bosonic}_{\pm,n;r}$ associated to \req{Weq}.

While the twisted form $\varphi^{bosonic}_{\pm,n}$ may directly be derived from the bosonic string disk amplitude for the construction of the  form $\widetilde{\varphi}^{Bosonic}_{\pm, n;r}$ we apply the embedding of disk fields onto the sphere following  \cite{MS}. In the  latter work we have introduced a map which embeds the field content of the open superstring sector onto a specific  closed superstring sector. This map has then  successively been applied to construct  the twisted form $\widetilde{\varphi}^{EYM}_{\pm;n;r}$.
In similar manner  we want to apply this map to  embed  the field content of a bosonic open string sector onto a closed string sector on the sphere and then extract the bosonic twisted form 
$\widetilde{\varphi}^{Bosonic}_{\pm,n;r}$. We begin with the embedding of the open string fields on the disk onto
the sphere as: 
\begin{equation}
   \begin{aligned}
       X^\mu(x) & \longmapsto X^\mu(z)+\widetilde{X}^\mu(\ov{z}) \ \Rightarrow \ \begin{cases}  & \partial X \mapsto \partial X \\
                                                                                    & k \cdot X \mapsto k \cdot (X+\widetilde{X})\ . \end{cases} 
         \end{aligned} \label{emb}
\end{equation}
Then, the bosonic open string vertex operators on the sphere are mapped according to \cite{MS}: 
\begin{equation}
   \begin{aligned}
  V_o(\varepsilon_i,k_i,x_i) &\longmapsto V_o(\varepsilon_i,k_i,z_i)\ e^{ik_i \widetilde{X}(\ov{z}_i)}\ \tilde J^{c_i}(\bar z_i)\ ,\ \ &i=1,\ldots,n\ ,\\
  V_c(\varepsilon_s,q_s,z_{n+s},\bar{z}_{n+s}) &\longmapsto V_c=  V_o(\varepsilon_s,q_s,z_{n+s}) V_o(\tilde{\varepsilon}_s,\tilde{q}_s,\bar{z}_{n+s}) & s=1,\ldots,r\ .
    \end{aligned}\label{MAP}
\end{equation}
Applying the map \req{MAP}  on a generic disk correlator involving $n$ open and $r$ closed bosonic strings  yields the following sphere correlator: 
\begin{equation}
   \begin{aligned}
  \Big\langle \prod\limits_{i=1}^n V_o(\varepsilon_i,k_i,z_i)& \prod\limits_{s=1}^r V_c(\varepsilon_s,q_s,z_{n+s},\bar{z}_{n+s}) \Big\rangle_{D_2}  \mapsto \langle \tilde J^{c_1}(\bar z_1)\ldots \tilde J^{c_n}(\bar z_n)\rangle_{S^2} \\
   &\times \Big\langle \prod\limits_{i=1}^n V_o(\varepsilon_i,k_i,z_i) e^{ik_i \tilde{X}(\bar{z}_i)} \prod\limits_{s=1}^r V_c(\varepsilon_s,q_s,z_{n+s},\bar{z}_{n+s}) \Big\rangle_{S_2}\ .
    \end{aligned}\label{amp}
\end{equation}
The factorization  property on the sphere allows to write the correlator \req{amp} as the product of holomorphic and anti--holomorphic contractions
\begin{equation}
   \begin{aligned}
\Ac_{embededd}(n;r)&= \underbrace{\Big\langle   \prod\limits_{i=1}^n V_o(\varepsilon_i,k_i,z_i)  \prod\limits_{s=1}^r V_o(\varepsilon_s,q_s,z_{n+s})  \Big\rangle_{S_2}}_{holomorphic}   \\
&\times      \underbrace{ \mathcal{C}_n\{\ov z_i\}\; \Big\langle  \prod\limits_{s=1}^r V_o(\tilde{\varepsilon}_s,\tilde{q}_s,\bar{z}_{n+s}) \prod\limits_{i=1}^n  e^{ik_i \tilde{X}(\bar{z}_i)} \Big\rangle_{S_2}}_{anti-holomorphic},
  \end{aligned} \label{ampemb}
\end{equation}
with the gauge current part 
\be
 \mathcal{C}_n(\{\ov z_i\}) = \frac{1}{(\bar z_1- \bar z_2)(\bar z_2-\bar z_3)\cdot\ldots\cdot(\bar z_n-\bar z_1)}\ ,\label{TWISTN1}
\end{equation}
referring to the single trace term $\Tr(T^{c_1}\ldots T^{c_n})$. The latter is extracted from the full gauge correlator $\langle \tilde J^{c_1}(\bar z_1)\ldots \tilde J^{c_n}(\bar z_n)\rangle_{S^2}$ by projection \cite{MS}. As  customary in \req{ampemb} the holomorphic part comprises  the contractions of $n+r$ bosonic open strings which  give rise to  the twisted form  $\varphi^{bosonic}_{\pm,n+r}$ given in \req{listT} and \cite{M1}. On the other hand,  the anti--holomorphic part of \req{ampemb} can be borrowed to
define $\widetilde{\varphi}^{Bosonic}_{\pm,n;r}$ as follows:
$$ \mathcal{C}_n\{\ov z_i\}\ \Big\langle  \prod\limits_{s=1}^r  V_o(\tilde{\varepsilon}_s,\tilde{q}_s,\bar{z}_{n+s})  \prod\limits_{i=1}^n  e^{ik_i \tilde{X}(\bar{z}_i)} \Big\rangle_{S_2}\times \ov{KN}^{-1}\;\Bigg|_{\bar{z_i} \mapsto z_i}=:\widetilde{\varphi}^{Bosonic}_{\pm,n;r}\ .$$
Note, that the above step involves a map  from anti--holomorphic  coordinates $\bar z_i$ to holomorphic ones $z_i$ together with the removal of the (anti--holomorphic) Koba--Nielsen factor $\ov{KN}$~\cite{MS}.
To this end, we find: 
\begin{align}
 \widetilde{\varphi}^{Bosonic}_{\pm,n;r}&= d\mu_{n+r} \;  \lf(\pm\frac{1}{\alpha'}\ri)^{\floor{\frac{n+r-2}{2}}} \mathcal{C}_n\{z_i\}\int \prod\limits_{i=1}^{r} d \theta_i d\bar{\theta_i}\; \nonumber\\ 
 & \times \exp  \Bigg\{ \pm2\alpha'\, \sum\limits_{i\in \Rc}\theta_i\bar\theta_i \, \Bigg(\sum\limits_{j\in\Nc } \frac{\varepsilon_i \cdot p_j}{z_i-z_j} + \sum\limits_{j\in\Rc \atop j \neq i} \frac{\varepsilon_i \cdot p_j}{z_i-z_j}\Bigg)+\sum\limits_{j\neq i \atop j,i \in \Rc}\frac{\theta_i \bar{\theta}_i \theta_j \bar{\theta}_j \varepsilon_i \cdot \varepsilon_j}{(z_i-z_j)^2}\Bigg\}\ .
\label{embW} 
\end{align}
Finally, we take the $\alpha' \rightarrow \infty$ limit as
\begin{equation}
\begin{aligned}
& \lim\limits_{\alpha' \rightarrow \infty} \widetilde{\varphi}^{Bosonic}_{\pm,n;r}=  \mathcal{C} _n\ \prod\limits_{i \in \Rc} \Bigg(\sum\limits_{j \in \Nc} \frac{\varepsilon_i \cdot p_j}{z_{ij}} + \sum\limits_{j \in \Rc \atop j \neq i} \frac{\varepsilon_i \cdot p_j}{z_{ij}}\Bigg) \ d\mu_{n+r} \equiv\mathcal{C}_n\ \widetilde{W}_{\underbrace{11...1}_{r}}\ d\mu_{n+r} \ ,
\end{aligned} \label{emblim}
\end{equation}
which comprises   the color form $\mathcal{C}_n$ (over the set of legs $\Nc$) and the function $\widetilde{W}_{\underbrace{11...1}_{r}}$ referring to the set $\Rc$ of legs and introduced in \req{Weq} .

\subsect{New theories constructed from the twisted form $\widetilde{\varphi}^{Bosonic}$}

In this subsection we display a set of theories whose amplitudes can be constructed from the intersection number \req{intersection} involving the  twisted form
 \req{phiB} or  our newly constructed form \req{embW}. In fact, the amplitudes are inferred from \req{alphlim} in the $\ap\ra\infty$ limit where \req{phiB} and \req{emblim}
become relevant.
The following  three twisted forms stemming from bosonic string theory are available:
\begin{itemize}
\item  $\varphi^{bosonic}_{\pm,n}$ arises from the full bosonic string amplitude and includes different orders in $\alpha'$, cf. \req{listT} and Table \ref{t1}. It does not directly correspond to a CHY representation.
\item  $\varphi^{Bosonic}_{\pm,n}$  is the $\alpha' \rightarrow \infty$ limit of $\varphi^{bosonic}_{\pm,n}$, cf. \req{phiB}. It  corresponds to the function $W_{\underbrace{11...1}_{n}}$ given in \req{listT2} in the CHY formulation. 
\item  $\widetilde{\varphi}^{Bosonic}_{\pm,n;r}$  is the embedded twisted $n+r$--form \req{embW} related to $\varphi^{bosonic}_{\pm,n}$. It  corresponds to the function $\widetilde{W}_{\underbrace{11...1}_{r}}$ given in \req{Weq} in the CHY formulation times a color $n$--form $\mathcal{C}_n$. 
\end{itemize}
 Building  up on the previous results we have the following table of theories and their different representations. 
\begin{table}[H]
\hskip-0.75cm%\centering
\begin{tabular}{|c| c c c c|} 
 \hline
&&&&\\
  Theory & $\varphi_-$ & $\varphi_+$ & CHY representation & Amplitude \\ [2ex] 
 \hline\hline
 $(DF)^2$ & $\varphi^{color}_{-,n}$ & $\varphi^{Bosonic}_{+,n}$ & $ \mathcal{C}_n\; W_{\underbrace{11...1}_{n}}  $ & \shortstack{ $n$ gluon \\ (higher derivative)}  \\  [2ex]  
 \hline
 Conformal Gravity (CG) & $\varphi^{gauge}_{-,n}$  & $\varphi^{Bosonic}_{+,n}$ & $  \pf' \psi_n \; W_{\underbrace{11...1}_{n}} $ &  $n$ graviton (Weyl)  \\   [2ex] 
 \hline
$(Weyl)^3 \,\,or \,\,R^3$   & $\varphi^{Bosonic}_{-,n}$  & $\varphi^{Bosonic}_{+,n}$ & $   \Bigg(W_{\underbrace{11...1}_{n}}\Bigg)^2 $ & \shortstack{ $n$ graviton \\ (higher derivative)}  \\   [2ex] 
 \hline
  $Weyl$--$YM$ & $\widetilde{\varphi}^{Bosonic}_{-,n;r}$  & ${\varphi}^{gauge}_{+,n+r}$ & $ \mathcal{C}_n\; \widetilde{W}_{\underbrace{11...1}_{r}}\; \pf' \psi_{n+r} $ & \shortstack{$r$ graviton (Weyl)\\ $n$ gluons } \\   [2ex] 
  \hline 
 $(DF)^2$--Photon & $\varphi^{scalar}_{-,n}$ & $\varphi^{Bosonic}_{+,n}$ &  $ (\pf' A_n)^2\; W_{\underbrace{11...1}_{n}}$ & $n$ higher derivative photon  \\ [2ex] 
 \hline
 $(DF)^2+\phi^3$& $\tilde{\varphi}^{Bosonic}_{-,n;r}$ & $\varphi^{color}_{+,n+r}$ &  $ \mathcal{C}_n\;\widetilde{W}_{\underbrace{11...1}_{r}}\;  \mathcal{C}_{n+r} $ & \shortstack{$r$ gluons (higher derivative) \\ $n$ scalars }  \\ [2ex] 
 \hline
\end{tabular}
\captionof{table}{Theories described by twisted intersection forms  ${\varphi}^{Bosonic}_{\pm,n}$ and $\tilde{\varphi}^{Bosonic}_{\pm,n;r}$.} \label{tt4}
\end{table}
\noindent 
Notice, that the construction of Weyl--YM is similar to the EYM amplitude \req{Amp EYM}. For the latter we  used the open and closed superstring together with the underlying embedding formalism to construct $\tilde \varphi^{EYM}_{-,n;r}$, while  $\tilde \varphi^{Bosonic}_{-,n;r}$ is derived from \req{MAP} involving the open and closed bosonic string. In the above table we have added the paring of our newly constructed twisted form \req{embW}
and the color form $\varphi^{color}_{\pm,n+r}$. In the $\alpha' \rightarrow \infty$ limit we obtain the corresponding CHY integrand from the twisted intersection \req{intersection}: 
\begin{equation}
\lim_{\alpha' \rightarrow \infty}\langle \tilde{\varphi}^{Bosonic}_{-,n;r}, \varphi^{color}_{+,n+r} \rangle_{\omega} \simeq  \mathcal{C}_n\;\widetilde{W}_{\underbrace{11...1}_{r}}\;  \mathcal{C}_{n+r} \ .\label{DFS}
\end{equation} 
We can see that the  resulting CHY representation  in (\ref{DFS}) corresponds to $(DF)^2+\phi^3$ theory.  
Finally, let us review some of the basics of the vector theories appearing in Table \ref{tt4}.

\begin{itemize}

\item Higher derivative gauge theory $(DF)^2$

In the literature there is a variety of higher derivative gauge theories denoted by $(DF)^2$. 
Here we use the following definition from \cite{Joh1}:
\begin{equation}
\mathcal{L}_{(DF)^2}=\frac{1}{2}(D_{\mu}F^{a\,\mu\nu})^2-\frac{g}{3}\;F^3+\frac{1}{2}(D_{\mu} \varphi^a)^2+\frac{g}{2} \; C^{\alpha \, ab} \varphi^{\alpha} F^{a}_{\mu\nu}F^{b \,\mu\nu} + \frac{g}{3!}\; d^{\alpha \beta \gamma} \varphi^{\alpha} \varphi^{\beta} \varphi^{\gamma}\ .
\label{LDF}
\end{equation}
In particular, we have the cubic higher order gauge interaction and gauge covariant derivative 
\begin{equation}
\begin{aligned}
 F^3&=f^{abc} F^{a \, \mu}_{\nu} F^{b \, \nu}_{\gamma} F^{c \, \gamma}_{\mu}\ ,\\
 D_{\mu} \varphi^\alpha&=\partial_{\mu} \varphi^{\alpha}-ig (T^a_R)^{\alpha \beta} A_{\mu}^a \, \varphi^{\beta}
 \end{aligned}
\end{equation}
respectively, with $\varphi$  in a real representation $R$ of the gauge group and real generators of the gauge group  $(T^a_R)^{\alpha \beta}$. This theory has six propagating degrees of freedom. One accounting for the "auxiliary" scalar field $\varphi$, which is not considered in the scattering amplitude under consideration. In addition, we have five degrees of freedom from the higher derivative gauge theory which includes a negative norm state (i.e. ghost).
The amplitude of this theory, which describes the scattering of $n$ higher derivative spin--one vectors,
is given by the intersection number \req{intersection} as
\begin{equation}
\begin{aligned}
 &\Ac_{(DF)^2}(n)  =\lim_{\ap\ra\infty}\langle\varphi_{-,n}^{color},\varphi_{+,n}^{Bosonic}\rangle_\omega=\int\limits_{\mathcal{M}_{0,n}} d\mu_n\ \prod_{k=2}^{n-2}\ \delta(f_k) \,\, \mathcal{C}_n\,\, W_{\underbrace{11...1}_{n}} \ ,
     \end{aligned}    \label{CHY DF}
     \end{equation}
 leading to the correct CHY formulation.

\item $(DF)^2-{Photon}$

Next, the $(DF)^2-{\rm Photon}$  theory involves the higher derivative $U(1)$  gauge theory and the Einstein gravity background. 
The relevant Lagrangian is given by: 
\begin{equation}
\begin{aligned}
 \frac{1}{\sqrt{-g}}\mathcal{L}_{(DF)^2_{Photon}}=& \frac{1}{2 \kappa^2} R+\frac{1}{4} (D_\mu F^{\alpha \beta})^2+\frac{1}{8} R F^2-\frac{1}{6}\kappa^2 D_\mu F_{\alpha \beta} D^\alpha F^{\mu \gamma} F_{\gamma \delta} F^{\delta \beta }\\
 &+ \frac{1}{48} \kappa^2 (D_\mu F^{\alpha \beta})^2 F^2+\mathcal{O}(\kappa^4)\ .
\end{aligned}\label{ActionDFP}
\end{equation}
The amplitude $\Ac_{(DF)^2_{Photon}}(n)$ involving $n$ photons 
can be computed by the intersection number \req{intersection}. The $\ap\ra\infty$ limit \req{alphlim} of the latter yields
\begin{equation}
\Ac_{(DF)^2_{Photon}}(n)  =\lim_{\ap\ra\infty}\lng\varphi^{scalar}_{-,n}, \varphi^{Bosonic}_{+,n} \rangle_{\omega}=\int\limits_{\mathcal{M}_{0,n}} d\mu_n\ \prod_{k=2}^{n-2} \delta(f_k) \,\, W_{\underbrace{11...1}_{n}}\; (\pf' A_n)^2\ ,
     \label{CHY DFPh}
     \end{equation}
and reproduces the  CHY representation of the amplitude $\Ac_{(DF)^2_{Photon}}(n)$ \citep{Aze}. 
In \req{CHY DFPh} we have  the submatrix $A_n$ defined in (\ref{psi1}) and the function $W_{\underbrace{11...1}_{n}}$ is given in (\ref{phiB}).

\end{itemize}

\sect{New and old double copies}
\label{NDC}

One of the most intriguing  features of double copy relations is the connection between spin--one and spin--two theories and their symmetries. Likewise, the relation between gauge invariance for spin--one theories and diffeomorphism invariance associated to spin--two theories. 
It has been argued in \cite{Joh1} that taking two spin--one theories which satisfy color--kinematics duality and double copy them  results in  a spin--two theory which is invariant under linearized diffeomorphism. This important feature requires further investigation and understanding for the massive case which is notable challenging \cite{MCK}. In that case one needs to clarify the role of  CK duality and the corresponding KK--BCJ amplitude relations. Then, with these informations one may  construct massive spin--one theories which satisfy this duality and check the corresponding double copied theory against diffeomorphism invariance for massive spin--two theories, e.g. dRGT gravity.

In intersection theory double copies can  simply be constructed by  pairing two theories whose description in terms of intersection numbers \req{intersection} comprise a color--form $\varphi^{color}\equiv PT(a)$. The latter constitutes  an orthonormal basis, i.e.
\be\label{ONB}
\vev{\Phi_{+,a},\Phi_{-,b}^\vee}_\om=\delta_{ab}\ \ \ ,\ \ \ \Phi_{+,a}=PT(a)\ ,
\ee
and this fact allows us to simply "glue" two different theories both containing such a color--form $\varphi^{color}$.
More concretely,  for two theories $T_1$ and $T_2$ given by the intersection numbers
\be\label{Theories}
T_1= \langle \varphi^1_+ ,\varphi^{color} \rangle \qquad \qquad T_2= \langle \varphi^2_- ,\varphi^{color}  \rangle
\ee
respectively, one schematically obtains for their double copy:
\be
T_1 \otimes T_2 =\langle \varphi^1_+ ,\varphi^2_- \rangle\ .
\end{equation}
More precisely, with the orthonormal basis \req{ONB} we have
\begin{align}
\varphi_+^1&=\sum_{a=1}^{(m-3)!}\vev{PT^\vee(a),\varphi_+^1}_\om\ PT(a)\ ,\nn\\
\varphi_-^2&=\sum_{b=1}^{(m-3)!}\vev{\varphi_-^2,PT(b)}_\om\ PT^\vee(b)\ ,\label{EXPF}
\end{align}
and with 
$$\vev{PT^\vee(a),\varphi_+^1}_\om=\sum_{b=1}^{(m-3)!} S[a|b]\ \vev{PT(b),\varphi_+^1}_\om\ ,$$ 
we consider the following manipulations leading to a double copy expression
\begin{align}
\vev{\varphi_+^1,\varphi^2_-}&=\sum_{a=1}^{(m-3)!}\vev{PT^\vee(a),\varphi_+^1}_\om\ \vev{\varphi_-^2,PT(a)}_\om\nn\\
&=\sum_{a,b=1}^{(m-3)!}\vev{\varphi_-^2,PT(a)}_\om \ S[a|b]\ \vev{PT(b),\varphi_+^1}_\om \nonumber\\
&=\sum_{a,b=1}^{(m-3)!} T_1(a)\ S[a|b]\ T_2(b)\ ,\label{DCE}
\end{align}
with the intersection form or KLT kernel $S[a|b]$ given in \req{Kernel}.
Therefore, the two theories \req{Theories} involving a color form in their twisted intersection forms   give rise to the double copy \req{DCE} denoted by $T_1 \otimes T_2$.

\subsection{Collections of some known double copies}
\renewcommand{\labelenumi}{(\roman{enumi})}

Here we compile a list of different double copy constructions from pairs of theories discussed before. The latter exhibit a color form $\varphi^{color}$  in their twisted intersection form \req{intersection}. In particular,  we shall discuss double copies for Born--Infeld and $(DF)^2_{Photon}$ theories.

\begin{enumerate}

\item  Special Galilean

We start with the special Galilean theory described by gluing two identical NSLM theories
\begin{align}
T_1(a)&=\Ac_{NLSM}(a)=\lim_{\ap\ra\infty}\vev{PT(a),\varphi^{scalar}_{+,n}}_\om\ ,\nn\\
T_2(b)&=\Ac_{NLSM}(b)=\lim_{\ap\ra\infty}\vev{\varphi^{scalar}_{-,n},PT(b)}_\om\ .
\end{align}
With \req{DCE} we construct the double copy $T_1 \otimes T_2$ of Galilean theory
\be
\Ac_{sGal}(n)=\lim_{\ap\ra\infty}\vev{\varphi^{scalar}_{-,n},\varphi^{scalar}_{+,n}}_\om
\ee
 describing the scattering of $n$ scalars with higher derivative interaction (\ref{sgal}), i.e.:
\be\label{DC1}
\Ac_{sGal}(n)=\sum_{a,b=1}^{(n-3)!}\Ac_{NLSM}(a)\ S[a|b]\ \Ac_{NLSM}(b)\ .
\ee
\item Einstein--Yang--Mills (EYM)

Secondly, we consider a double copy from gen.YMS and YM theories
\begin{align}
T_1(a)&=\Ac_{gen.YMS}(a)=\lim_{\ap\ra\infty}\vev{PT(a),\tilde\varphi^{EYM}_{+,n;r}}_\om\ ,\nn\\
T_2(b)&=\Ac_{YM}(b)=\vev{\varphi^{gauge}_{-,n+r},PT(b)}_\om
\end{align}
to construct the double copy $T_1 \otimes T_2$ of EYM amplitudes  \cite{MS}
\be
\Ac_{EYM}(n;r)=\lim_{\ap\ra\infty}\vev{\varphi^{gauge}_{-,n+r},\tilde\varphi^{EYM}_{+,n;r}}_\om \label{DCEYM}
\ee
describing the scattering of $n$ gluons and $r$ gravitons  \cite{Cachazo2014}
\be\label{DC2}
\Ac_{EYM}(n;r)=\sum_{a,b=1}^{(m-3)!}\Ac_{gen.YMS}(a)\ S[a|b]\ \Ac_{YM}(b)\ ,
\ee
with $m=n+r$.

\item Dirac--Born--Infeld (DBI)

Thirdly, we glue together YMS and NSLM theories
\begin{align}
T_1(a)&=\Ac_{YMS}(a)=\lim_{\ap\ra\infty}\vev{PT(a),\tilde{\varphi}^{EM}_{+,n;r}}_\om\ ,\nn\\
T_2(b)&=\Ac_{NSLM}(b)=\lim_{\ap\ra\infty}\vev{\varphi^{scalar}_{-,n+r},PT(b)}_\om
\end{align}
to construct the double copy $T_1 \otimes T_2$ of DBI amplitudes
\be
\Ac_{DBI}(n;r)=\lim_{\ap\ra\infty}\vev{\varphi^{scalar}_{-,n+r}, \tilde{\varphi}^{EM}_{+,n;r}}_\om
\ee
describing the scattering of $r$ gluons and $n$ scalars  \cite{Cachazo2014}:
\be\label{DC3}
\Ac_{DBI}(n;r)=\sum_{a,b=1}^{(m-3)!}\Ac_{YMS}(a)\ S[a|b]\ \Ac_{NLSM}(b)
\ee
\item Einstein--Maxwell (EM)

Furthermore,  EM can be written as double copy of the following two theories 
\begin{align}
T_1(a)&=\Ac_{YMS}(a)=\lim_{\ap\ra\infty}\vev{PT(a),\tilde{\varphi}^{EM}_{+,n;r}}_\om\ ,\nn\\
T_2(b)&=\Ac_{YM}(b)=\vev{\varphi^{gauge}_{-,n+r},PT(b)}_\om\ .
\end{align}
giving rise to the double copy $T_1 \otimes T_2$ of EM amplitudes
\be
\Ac_{EM}(n;r)=\lim_{\ap\ra\infty}\vev{\varphi^{gauge}_{-,n+r}, \tilde{\varphi}^{EM}_{+,n;r}}_\om 
\ee
involving of $r$ gravitons and $n$ photons \cite{Cachazo2014}: 
\be\label{DC4}
\Ac_{EM}(n;r)=\sum_{a,b=1}^{(m-3)!}\Ac_{YMS}(a)\ S[a|b]\ \Ac_{YM}(b)\ .
\ee
\item Born--Infeld (BI) theory

 The amplitudes of Born--Infeld theory  can be written as double copy of the following two theories:
\begin{align}
T_1(a)&=\Ac_{NLSM}(a)=\lim_{\ap\ra\infty}\vev{PT(a),\varphi^{scalar}_{+,n}}_\om\ ,\nn\\
T_2(b)&=\Ac_{YM}(b)=\vev{\varphi^{gauge}_{-,n},PT(b)}_\om\ .
\end{align}
Then,  the double copy $T_1 \otimes T_2$ yields the BI amplitudes 
\be
\Ac_{BI}(n)=\lim_{\ap\ra\infty}\vev{\varphi^{gauge}_{-,n}, \varphi^{scalar}_{+,n}}_\om 
\ee 
accounting for the scattering of $n$ gluons:
 \be\label{DC7}
\Ac_{BI}(n)=\sum_{a,b=1}^{(n-3)!}\Ac_{NLSM}(a)\ S[a|b]\ \Ac_{YM}(b)\ .
\ee

\item Extended Dirac Born--Infeld (ext.DBI) theory

In addition to BI and DBI the amplitudes of ext.DBI theory can be written as double copy of the following two theories:
\begin{align}
T_1(a)&=\Ac_{NLSM}(a)=\lim_{\ap\ra\infty}\vev{PT(a),\varphi^{scalar}_{+,n+r}}_\om\ ,\nn\\
T_2(b)&=\Ac_{gen.YMS}(b)=\vev{\tilde{\varphi}^{EYM}_{-,n;r},PT(b)}_\om\ .
\end{align}
Then,  the double copy $T_1 \otimes T_2$ yields the ext.DBI amplitudes 
\be
\Ac_{ext.DBI}(n,r)=\lim_{\ap\ra\infty}\vev{\tilde{\varphi}^{EYM}_{-,n;r}, \varphi^{scalar}_{+,n+r}}_\om 
\ee 
accounting for the scattering of $r$ gluons and $n$ higher derivative scalars as:
 \be\label{DC5}
\Ac_{ext.DBI}(n,r)=\sum_{a,b=1}^{(m-3)!}\Ac_{NLSM}(a)\ S[a|b]\ \Ac_{gen.YMS}(b)\ .
\ee

\item $(DF)^2_{Photon}$ theory

Finally, the amplitudes \req{CHY DFPh} of $(DF)^2_{Photon}$ theory can be written by double copying the following two theories:
\begin{align}
T_1(a)&=\Ac_{NLSM}(a)=\lim_{\ap\ra\infty}\vev{PT(a),\varphi^{scalar}_{+,n}}_\om\ ,\nn\\
T_2(b)&=\Ac_{(DF)^2}(b)=\vev{\varphi^{Bosonic}_{-,n},PT(b)}_\om\ .
\end{align}
The double copy $T_1 \otimes T_2$ for the amplitudes 
\be
\Ac_{(DF)^2_{Photon}}(n)=\lim_{\ap\ra\infty}\vev{\varphi^{Bosonic}_{-,n}, \varphi^{scalar}_{+,n}}_\om 
\ee 
involving $n$ higher derivative photons gives:
\be\label{DC6}
\Ac_{(DF)^2_{Photon}}(n)=\sum_{a,b=1}^{(n-3)!}\Ac_{NLSM}(a)\ S[a|b]\ \Ac_{(DF)^2}(b)\ .
\ee
Note, that this double copy involves Einstein gravity and higher--derivative photon terms, cf. Eq.  \req{ActionDFP}.
\end{enumerate}

\noindent
Finally, let us collect all seven double copies $(i)$--$(vii)$ in the following table.
%\vskip-0.5cm
\begingroup
\renewcommand{\arraystretch}{1.3} % Default value: 1
\begin{table}[H]
\centering
\begin{tabular}{|c| c c c |} 
 \hline
&&&\\
  Theory & $T_1$ & $T_2$ & CHY representation \\ [2ex] 
 \hline\hline
 sGal &   NSLM &  NSLM & $(\pf' A_n)^4$  \\ [2ex] 
 \hline
 EYM &  gen.YMS &  YM & ${\cal C}_n \pf\Psi_{S_r}\,\pf'\psi_{n+r}$     \\   [2ex] 
  \hline 
DBI &  YMS &  NSLM & $  \pf X_n\, \pf' \Psi_{S_{r:n}}\;(\pf' A_{n+r})^2 $ \\   [2ex]
 \hline 
EM &  YMS &  YM& $  \pf X_n\, \pf' \Psi_{S_{r:n}}\;\pf'\psi_{n+r} $ \\   [2ex]
 \hline 
  BI & NSLM  & YM & $  (\pf' A_n)^2\; \pf' \psi_{n} $\\   [2ex] 
  \hline 
ext.DBI & NSLM &  YM & $ \mathcal{C}_n  \;\pf \Psi_{S_r} \; (\pf' A_{n+r})^2 $ \\   [2ex]
 \hline 
  $DF^2$-photon & NSLM &  $(DF)^2$ &  $ (\pf' A_n)^2\; W_{\underbrace{11...1}_{n}}$   \\ [2ex] 
 \hline 
\end{tabular}
\captionof{table}{Double copies $T_1\otimes T_2$ through twisted form description \req{Theories}.} \label{tnew}
\end{table}
\endgroup

\subsection{Further double copies}

As  discussed in \cite{Aze1,Bros} double coping spin--one theories, which satisfy color--kinematics duality, gives rise   to the existence of a diffeomorphism invariant spin--two theory in the resulting product. This is apparent both at the level of the amplitudes and the Lagrangian. Our amplitudes  are computed by  intersection numbers \req{intersection} supplemented by  a pair of twisted forms. If the latter comprise a color form  $\varphi^{color}$, i.e. we  have the representations \req{Theories}, then generically  CK duality is furnished for this theory, cf. Subsection \ref{AmpsRel}. Therefore, those theories \req{Theories} are  candidates for building blocks of  double copies \req{DCE} or likewise for new spin--two theories. In the following Table \ref{t5} we
compile a list of all previously known and newly constructed double copies   from theories containing a color form $\varphi^{color}$. 
\begin{table}[H]
\renewcommand{\thetable}{\arabic{table}a}
\hskip-0.25cm%\centering
{\small
\begin{tabular}{|c| c  c c c|} 
 \hline
 &&&&\\
  Theory $\otimes$ &  $(DF)^2 $  & YM & NLSM & YM+$(DF)^2$ \\ [2ex] 
 \hline
   $(DF)^2$ & Weyl$^3$ or $R^3$\, Gravity  & Conformal Gravity & $(DF)^2$-Photon  & $R^3$-Weyl  \\  [2ex]  
 \hline
 YM  & Conformal Gravity  &  GR   &Born Infeld & Einstein--Weyl \\   [2ex] 
 \hline
  NLSM & $(DF)^2$-Photon    & Born-Infeld  & Special Galilean &  BI-$(DF)^2$ Photon \\   [2ex] 
 \hline
  YM+$(DF)^2$ & $R^3$-Weyl   & Einstein--Weyl & BI-$(DF)^2$ Photon  &   HD gravity \\   [2ex] 
 \hline
\end{tabular}
\captionof{table}{Table of different double copies} \label{t5}}
\end{table}
\noindent E.g. the double copy of $(DF)^2$ with itself gives rise to the amplitudes of six--derivative gravity  $R^3$ originating from the bosonic ambitwistor string \cite{Aze}. Note, in contrast  double copying  $F^3$ amplitudes leads to the amplitudes produced by Einstein gravity coupled to a dilaton field $\phi$, and deformed by operators of the form $\phi R^2$ and $R^3$ \cite{Broedel:2012rc}. 
On the other hand, the conformal supergravity (CSG) amplitudes (of the non--minimal Berkovits--Witten type theory \cite{Ber04}) from Table \ref{tt4} follow from double copying $(DF)^2$ and SYM theory \cite{Joh1}.
From Table \ref{t5} it can be evidenced, that there are new double copies of theories which can be constructed only by using the property  that the color form provides CK duality and can be used as integral part to double copy two different theories. Besides, there is a close relation between string theory  and the content of Table \ref{t5} by means of looking at the inherent double copy of the closed string in terms of open strings cf. \citep{klt,Aze1,Siegel}. Moreover, we find a double copy for higher derivative (HD) gravity and bigravity to be discussed in Subsections \ref{SHD} and \ref{PMBI}, respectively.

Furthermore, in the next Table \ref{t6} the four double copies \req{DC2}, \req{DC3}, \req{DC4} and \req{DC5} are tabulated among some new  constructions like EM squared.
\begin{table}[H]
\addtocounter{table}{-1}
  \renewcommand{\thetable}{\arabic{table}b}
\centering
{\small
\begin{tabular}{|c| c  c c c|} 
 \hline
 &&&&\\
  Theory $\otimes$ &  YMS  & gen.YMS & NLSM & YM \\  [2ex] 
 \hline
   YMS   & EM$^2$  & EM $+$ YMS & DBI & EM  \\  [2ex]  
 \hline
 gen.YMS  & EM $+$ YMS  &  EYMS  & ext.DBI & EYM \\   [2ex] 
 \hline
\end{tabular}
\captionof{table}{Table of different double copies} \label{t6}}
\end{table}
\noindent
We use the property, that cubic bi--adjoint scalar theory $\phi^3$ acts as identity element if double copied with another theory or itself \cite{Chi:2021mio}. For example the double copy construction (\ref{DCEYM}) of EYM amounts to:
\begin{equation}
\text{gen.YMS} \otimes \text{YM}= [YM+\phi^3] \otimes YM= (YM \otimes YM) + YM= GR+YM=\text{EYM}\ . 
\end{equation}
Furthermore, in Table \ref{t6} we have:
\begin{align}
\text{gen.YMS} \otimes \text{YMS}&= [YM+\phi^3] \otimes YMS= (YM \otimes YMS) + YMS= \text{EM}+\text{YMS}\ .
\end{align}
On the other hand, double copying two copies of gen.YMS amplitudes has been argued to produce 
amplitudes in Einstein--Yang--Mills--Scalar (EYMS) theory. The latter simply follows from compactification of EYM theory \cite{chy,Chiodaroli:2014xia,Cachazo2014}.
This result can also been anticipated from the decomposition:
\begin{align}
\text{gen.YMS} \otimes \text{gen.YMS} &= \text{gen.YMS} \otimes [YM+\phi^3]= EYM + \text{gen.YMS} \ .
\end{align}

\subsection{Double copies of spin--two theories}

In Table \ref{t5} we have gathered various spin--two theories including GR. While 
the latter is well--known the other theories may not be familiar. Therefore, for those we shall discuss their Lagrangians and relevant amplitudes. Actually,  we need to emphasize that for some of these theories the exact form of their Lagrangian is not known or  different names for them are used by various authors.

\subsubsection{Conformal Gravity}

We start with conformal gravity (CG). This name is  associated to different theories. The simplest definition corresponds to the $({\rm Weyl})^2$ action (pure $(\text{Weyl})^2$ conformal gravity), i.e.
\be\label{WeylG}
\mathcal{L}_{CG}=\kappa_W^{-2}\; \sqrt{-g}\ (W_{\mu\nu\alpha\beta})^2\ ,
\ee
with the  coupling constant $\kappa_W$ and  $W_{\mu\nu\alpha\beta}$ the Weyl tensor
\be
W_{\al\bet \gamma \delta}=R_{\al\bet\gamma\delta}-\h\;(R_{\al  \gamma}g_{\beta\delta}-R_{\al \delta}g_{\beta \gamma}+R_{\beta \delta}g_{\al \gamma}-R_{\beta  \gamma}g_{\al \delta})+\fc{1}{6}\;
(g_{\al  \gamma}g_{\beta \delta}-g_{\al \delta}g_{\beta  \gamma})\; R\ ,
\ee
which has the same symmetry  properties (w.r.t. its indices) than the Riemann tensor $R_{\al\bet \gamma \delta}$. 
In addition to the square of the Riemann tensor the square of the Weyl tensor is the second independent quadratic curvature invariant. Actually, we have the following relation:
\be\label{WeylGB}
W_{\mu\nu\rho\sigma}W^{\mu\nu\rho\sigma}=R_{GB}+2\;\lf(R_{\mu\nu}R^{\mu\nu}-\fc{1}{3}\;R^2\ri)\ ,
\ee
with the  Gauss--Bonnet term
\be\label{GaussB}
R_{GB}=R_{\mu\nu\rho\sigma}R^{\mu\nu\rho\sigma}-4\;R_{\mu\nu}R^{\mu\nu}+R^2\ ,
\ee
which for $d=4$ reduces to a  topological surface term. 
The latter can be added to the action without changing the (classical) theory in a spacetime which is asymptotically Minkowski.
Therefore, in $d=4$ the Lagrangian \req{WeylG} can be written in terms of the Riemann scalar $R$ and Ricci tensor $R_{\mu\nu}$ as: 
\begin{equation}
\begin{aligned}
& \mathcal{L}_{CG}= \kappa_W^{-2}\;  \sqrt{-g}\  \lf(\frac{1}{3}R^2-R^{\mu\nu}R_{\mu\nu}\ri)\ . \label{cg}
\end{aligned}   
 \end{equation}
According to Table \ref{t5}  the scattering amplitude $\Ac_{CG}(n)$ of this theory  involving  $n$ spin--two particles can be represented as a double copy in agreement with the CHY representation \cite{Aze}
\begin{equation}
\Ac_{CG}(n)\sim \lim_{\alpha' \rightarrow \infty}\langle \varphi^{gauge}_{-,n}, \varphi^{Bosonic}_{+,n} \rangle_{\omega}=\int\limits_{\mathcal{M}_{0,n}} d\mu_n \prod_{k=2}^{n-2} \delta(f_k) \,\, \pf' \psi_{n}\,\, W_{\underbrace{11...1}_{n}}\ ,
\label{cgdc}
\end{equation}
with the object\footnote{We should emphasize that the function $W_{\underbrace{11...1}_{r}}$ in the CHY representation \req{cgdc} does not refer to the Weyl tensor.} \req{listT2}
\begin{equation}
 W_{\underbrace{11...1}_{r}}= \widetilde{W}_{\underbrace{11...1}_{r}} \Big|_{\Nc=0}= \prod\limits_{i=1}^r \Bigg( \sum\limits_{j=1 \atop j \neq i}^r \frac{\varepsilon_i \cdot p_j}{z_{ij}}\Bigg)\ ,
 \label{Wn}
\end{equation}
derived from \req{Weq} for $n=0$. CG theory %\footnote{{\color{red}We should emphasize here as much as it has been hinted to be related to confromal gravity (the naming is a clear evidence) it has not been shown to exactly correspond to the confromal gravity in the from of (\ref{cg}). What has been shown for 4 point amplitude is that it corresponds to Berkovits-Witten super conformal gravity \cite{Ber04}. However, looking at the bosonic part of the theory  it will correspond to  Weyl+axion \cite{Joh1}.}}%
propagates  six degrees of freedom (packaged into the metric $g_{\mu\nu}$) which contain a massless spin--two (two degrees of freedom), a massless spin--one (two degrees) and a massless ghost spin--two field (two degrees) \cite{RIE}.

In (\ref{cgdc}) we have exhibited the double copy structure of CG \cite{Joh1}:
\begin{equation}\label{DCCG}
{\rm Conformal\ Gravity}=(DF)^2 \otimes YM 
\end{equation} 
in terms of intersection numbers \req{intersection}.
Note that both factors $(DF)^2$ and $YM$ have vectors $\tilde A_\mu$ and $A_\nu$, respectively, which 
tensor according to
\be\label{discuss1}
\tilde A_\mu \otimes A_\nu=g_{\mu\nu}\oplus B_{\mu\nu}\otimes \phi
\ee
giving rise to a graviton $g_{\mu\nu}$, an  antisymmetric two--form $B_{\mu\nu}$ and  a dilaton field $\phi$. Looking at the double copy \req{cgdc} we evidence the origin of the massless spin--one degrees of freedom as descending from the YM theory and the massless spin--two and ghost spin--two fields are stemming from $(DF)^2$.

\subsubsection{Einstein--Weyl gravity}\label{WEW}

Einstein--Weyl gravity is a modification of GR by adding the  square of the Weyl tensor \cite{RIE,Fer}:
\begin{equation}\label{EWaction}
\mathcal{L}_{EW}= \sqrt{-g}\; \lf(m^2\; R+\kappa_W^{-2}\; W_{\mu\nu\alpha\beta}^2\ri)\ .
\end{equation}

Above, the parameter $m$ relates to the string scale $m^2 \sim \alpha'^{-1}$ (likewise to the Planck scale $m\sim M_{\rm Planck}$). In fact, the mass parameter $m$ interpolates between two--derivative and four--derivative theories. This way by taking the limit $\alpha' \rightarrow \infty$ we reproduce pure $(\text{Weyl})^2$ conformal gravity. Similar to equation (\ref{masslim}) we have:
\begin{equation}\label{discuss2}
\lim_{\alpha' \rightarrow \infty} \mathcal{L}_{EW}(\alpha')\simeq\mathcal{L}_{W}\ , 
\end{equation}
On the other hand, for $\ap\ra0$ we end up at (non--pure) Einstein gravity. It can be shown, that the 
EW  action \req{EWaction} describes both massless and massive spin--two up to total derivatives. EW gravity has seven degrees of freedom accounting for two degrees from the standard massless spin--two graviton and an additional five (ghost) degrees for the massive spin--two field all packaged inside $g_{\mu\nu}$.
In fact, starting from an action for  a particular bimetric gravity with the two spin--two fields $g_{\mu\nu}$ and 
$f_{\mu\nu}$ with a mass term for the latter and eliminating $f_{\mu\nu}$ through  its equations of motion yields  \req{EWaction} \cite{Ferrara:2018wlb}. We shall return to this properties in Subsection \ref{PMBI}. 

For EW gravity scattering amplitudes  $\Ac_{EW}(n)$ involving $n$ gravitons  can be constructed through bosonic string amplitudes \cite{Aze1}. 
According to \req{discuss2}  in the limit $\ap\ra\infty$ the Einstein--Hilbert part of the EW theory decouples and therefore the CHY representation of this theory cannot be constructed by applying this  limit  at  the  intersection number \req{alphlim} of twisted forms. On the other hand, by using the bosonic string content one can write the amplitudes  of the full theory  as \cite{M1} (cf. also Table \ref{t1}):
\begin{equation}
\Ac_{EW}(n) =\lng\varphi^{gauge}_{+,n}, \varphi^{bosonic}_{-,n} \rangle_{\omega}\ .
    \label{CHY EW}
\end{equation}
The double copy structure of this theory is given by: 
\begin{equation}\label{DEW}
\textit{\rm Einstein--Weyl}=[YM+(DF)^2] \otimes YM \sim \langle \varphi^{gauge}_{-,n}, \varphi^{bosonic}_{+,n} \rangle_{\omega}\ .
\end{equation}
Note, that according to \req{masslim} and \req{discuss2} in the limit $\ap\ra\infty$ the double copy structure \req{DEW} will reduce to the  double copy of conformal gravity \req{DCCG}.
On the other hand, for $\ap\ra0$ we obtain the double copy for non--pure Einstein gravity including an anti--symmetric tensor and  a dilaton scalar, cf. also  \req{discuss1}.

Actually, the double copy structure \req{DEW} resembles the heterotic string. The amplitudes for the closed heterotic string also referred to as GR+$R^2$ can be calculated by using the KLT relations implementing the open bosonic string amplitude \req{OBS} together with the open superstring amplitude \req{OSS}. The action up to order $\ap^3$ was derived in \cite{Gross:1986mw}  
\begin{align}\label{GrossSloan}
S_{\rm{heterotic\ string}} & =-\fc{2}{\kappa^{2}}  
\int {\mathrm d}^d x \, \sqrt{g}\;  \left\{
R -\tfrac{4}{d-2}(\partial_{\mu}\phi)^2 - \frac{1}{12} H^2 \right.\\
&\qquad\qquad + \frac{\alpha'}{8} e^{-2\phi} \;\big( R_{\mu\nu\lambda\rho} R^{\mu\nu\lambda\rho} - 4 R_{\mu\nu} R^{\mu\nu}+R^2 \big)+ {\cal O}(\alpha'^3)  \Big\},\nonumber
\end{align}
where  $\phi$ represents the  dilaton and $H_{\mu\nu\rho}=3 \lf(\p_{[\mu}B_{\nu\rho]}+\tfrac{\ap}{4}\omega^{ab}_{[\mu}R^{ab}{}_{\nu\rho}\ri)$ is the field strength of the anti--symmetric $B$--field and the Chern--Simons form $\omega$. Note, that in \req{GrossSloan} the linear order in $\ap$ corresponds to the  Gauss--Bonnet term \req{GaussB}.
In the limit $\ap\ra0$ we recover standard GR  including an anti--symmetric tensor and  a dilaton scalar, cf. \req{discuss1}. Likewise, in this limit $\ap\ra0$ the double copy structure \req{DEW} boils down to
$YM\otimes YM$.

\subsubsection{$Weyl^3$ or $R^3$ gravity} 

Another spin-$2$ higher derivative theory can be constructed through the six derivative terms like $Weyl^3$ or $R^3$. Some speculation about the existence of $Weyl^3$ is made in \cite{Mason:2013}, where arguments on dimensional analysis are presented. The amplitude structure can be reproduced by the intersection number \req{intersection} involving a pair of the twisted forms \req{phiB}
 \be
\Ac_{Weyl^3}(n)  =\lim_{\alpha' \rightarrow \infty}\langle \varphi^{ Bosonic}_{-,n}, \varphi^{ Bosonic}_{+,n} 
\rangle_{\omega} \ ,  
     \end{equation}
leading to the  following CHY representation:
 \begin{equation}
\Ac_{Weyl^3}(n)   = \int\limits_{\mathcal{M}_{0,n}} d\mu_n\ \prod_{k=2}^{n-2} \delta(f_k) \,\, (W_{\underbrace{11...1}_{n}})^2 \ .
    \label{CHY W3}
\end{equation}
Looking at equation (\ref{CHY DF}) it is evident that this integrand is built from two sectors 
 accounting for $(DF)^2$. Therefore,  the twisted form \req{phiB} gives rise to  the double copy structure of the amplitude \req{CHY W3}, i.e.: 
 \begin{equation}
Weyl^3=(DF)^2 \otimes (DF)^2\ ,
\end{equation} 
or likewise \cite{Joh1}:
\begin{equation}
(DF)^2 \otimes (DF)^2 \sim (\nabla R)^2+R^3\ .
\end{equation}

\subsubsection{$Weyl^3$--$(DF)^2$}

Here we shall discuss a double copy of $(DF)^2$ with $(DF)^2+\phi^3$. More specifically, we shall discuss how the amplitude of $Weyl^3$--$(DF)^2$ theory may be described by the two twisted forms $\varphi^{Bosonic}_{\pm,n+r}$ and $\tilde{\varphi}^{Bosonic}_{\pm,n;r}$. The $Weyl^3$--$(DF)^2$ theory can be considered as an exotic cousin of EYM theory. Its CHY representation is assumed to take the following form: 
\begin{equation}
\begin{aligned}
 \Ac_{Weyl^3-(DF)^2}(n;r)&  =\lim\limits_{\alpha' \rightarrow \infty} \langle \varphi^{Bosonic}_{-,n+r}, \tilde{\varphi}^{Bosonic}_{+,n;r} \rangle_{\omega} =\int\limits_{\mathcal{M}_{0,n+r}} d\mu_{n+r}\ \prod_{k=2}^{n+r-2} \delta(f_k) \,\, \mathcal{C}_n \widetilde{W}_{\underbrace{11...1}_{r}} W_{\underbrace{11...1}_{n+r}}\ .
     \end{aligned}    \label{W3DF}
     \end{equation}
The formula \req{W3DF} is a conjecture, which presumably may be  proven  by ambitwistor construction \cite{Mason:2013}. However, this goes beyond the scope of this work and thus we write the $Weyl^3$-$(DF)^2$ theory in terms of the following double copy of theories:
\begin{equation}
 Weyl^3-(DF)^2=(DF)^2\otimes  [(DF)^2+\phi^3]=(DF)^2\otimes  (DF)^2+(DF)^2\ .
\end{equation}

\subsubsection{Summary}

In the following Table \ref{t7} we summarize  old and newly constructed double copies  and their pairs of twisted forms leading to spin--two theories.

%\vskip-0.5cm
\begingroup
\renewcommand{\arraystretch}{1.3} % Default value: 1
\begin{table}[H]
\centering
\begin{tabular}{|c| c c c |} 
 \hline
&&&\\
  Theory & $\varphi_-$ & $\varphi_+$ & CHY representation  \\ [2ex] 
 \hline\hline
 Einstein--Weyl & $\varphi^{gauge}_{-,n}$ & $\varphi^{bosonic}_{+,n}$ & ??   \\  [2ex]  
 \hline
 Conformal Gravity (CG) & $\varphi^{gauge}_{-,n}$  & $\varphi^{Bosonic}_{+,n}$ & $  \pf' \psi_n W_{\underbrace{11...1}_{n}} $   \\   [2ex] 
 \hline
 $Weyl^3$ & $\varphi^{Bosonic}_{-,n}$  & $\varphi^{Bosonic}_{+,n}$ & $   \Bigg(W_{\underbrace{11...1}_{n}}\Bigg)^2 $   \\   [2ex] 
 \hline
 $Weyl^3-DF^2$  & $\widetilde{\varphi}^{Bosonic}_{-,n;r}$  & $\varphi^{Bosonic}_{+,n+r}$  &  $\mathcal{C}_n \widetilde{W}_{\underbrace{11...1}_{r}} W_{\underbrace{11...1}_{n+r}}$  \\ [2ex] 
 \hline
 HD gravity &  $\varphi^{bosonic}_{-,n}$ & $\varphi^{bosonic}_{+,n}$ & $??$  \\ [2ex] 
 \hline
\end{tabular}
\captionof{table}{Old and new double copies for spin--two theories.} \label{t7}
\end{table}
\endgroup

\subsect{Double copy of $YM+(DF)^2$ and higher derivative gravity}\label{SHD}

In this section we look at the double copy of  $YM+(DF)^2$ theory and show that it can be related to higher derivative  (HD) gravity. Looking at the Table \ref{t5} we can write the double copy of the $YM+(DF)^2$ in terms of known spin--two theories in the following way:
\begin{equation}
\begin{aligned}
 [(DF)^2 + YM] &\otimes [(DF)^2 + YM] \\[2mm]
&\sim  \underbrace{( YM\otimes YM )}_{GR}+ \underbrace{( DF^2 \otimes DF^2 )}_{R^3}+ \underbrace{( DF^2 \otimes YM )}_{CG}+\underbrace{( YM \otimes DF^2 )}_{CG}\\[2mm]
&  \sim GR+CG_1+CG_2+R^3\sim GR+\widetilde{CG}+\mathcal{O}(R^3)\ . \label{DCYMDF}
\end{aligned}
\end{equation}
Since the amplitude of the $YM+(DF)^2$ is described by the intersection number \req{intersection} of $\varphi^{bosonic}_{\pm,n}$ and $\varphi^{color}_{\pm,n} $ we can write the amplitude of the resulting  double copied theory as: 
\be 
{\Ac}(n)=\langle \varphi^{bosonic}_{-,n}, \varphi^{bosonic}_{+,n}\rangle\ .  \label{AMPHD}
\ee
The twisted form $\varphi^{bosonic}_{\pm,n}$ stems  from the bosonic string and is  given in \req{listT}. 
The amplitude \req{AMPHD} only involves the metric $g_{\mu\nu}$ described by the higher derivative Lagrangian, which comprises  the conformal term+GR+$R^3$ corrections \req{DCYMDF}. Therefore, looking at the structure of the double copy in (\ref{DCYMDF})
we see that this amplitude corresponds to the interaction of $n$ spin--two fields (i.e.$\, g_{\mu\nu}$) with higher derivative interactions. Hence, it is worth looking at the theories known as HD gravity. The latter can be defined by the following Lagrangian \cite{Stelle:1977ry}:
 \begin{equation}
\begin{aligned}
  \mathcal{L}=& m_g^{2}\; \sqrt{g}\; \Bigg[\lambda_1 R(g)+\lambda_3 \;\lf(\fc{1}{3}R^2-R^{\mu\nu}R_{\mu\nu}\ri) \Bigg]+\mathcal{O}(R^3)\ ,\label{HDGR}
\end{aligned}
\end{equation}
with the following equation of motion
\begin{equation}\label{eqhd}
 B_{\mu\nu}+\fc{\lambda_1}{\lambda_3}\ G_{\mu\nu}+\mathcal{O}(R^3)=0\ ,
\end{equation}
where  $B_{\mu\nu}$ and $G_{\mu\nu}$ are the Bach and Einstein tensors, respectively. Some possible candidates for the $\mathcal{O}(R^3)$ operators in gravity can be found in \cite{Bueno:2016xff}.  
The Lagrangian in (\ref{HDGR}) describes GR  plus Weyl (conformal) interactions together with higher order curvature corrections $\mathcal{O}(R^3)$. Comparing this Lagrangian to the double copy structure in (\ref{DCYMDF}) we can see that the content and the type of the interactions are the same. 

Actually,  the amplitudes for the closed bosonic string also referred to as GR+$R^2$+$R^3$ can be calculated by using the KLT relations and the open bosonic string amplitudes \req{OBS}. The action up to order $\ap^3$ was derived in \cite{Metsaev:1986yb}  
\begin{align}\label{Metsaev}
S_{\substack{\rm{closed} \\ \rm{bosonic\ string}}} & =-\fc{2}{\kappa^{2}}  
\int {\mathrm d}^d x \, \sqrt{g}  \left\{
R -\tfrac{4}{d-2}(\partial_{\mu}\phi)^2 - \frac{1}{12} H^2 \right.\\
&\qquad\qquad + \frac{\alpha'}{4} e^{-2\phi} \;\big( R_{\mu\nu\lambda\rho} R^{\mu\nu\lambda\rho} - 4 R_{\mu\nu} R^{\mu\nu}+R^2 \big) \nonumber \\
&\qquad\qquad +\alpha'^2  e^{-4\phi}\;  \bigg( \frac{1}{16} R^{\mu \nu}{}_{\alpha\beta} R^{\alpha\beta}{}_{\lambda\rho} R^{\lambda\rho}{}_{\mu \nu}
- \frac{1}{12}R^{\mu\nu}{}_{\alpha\beta} R^{\nu\lambda}{}_{\beta\rho} R^{\lambda\mu}{}_{\rho \alpha} \bigg)+ {\cal O}(\alpha'^3)  \Big\},\nonumber
\end{align}
where  $\phi$ represents the  dilaton and $H_{\mu\nu\rho}=3 \p_{[\mu}B_{\nu\rho]}$ is the field strength of the anti--symmetric $B$--field. Note, that in \req{Metsaev} the linear order in $\ap$ corresponds to the  Gauss--Bonnet term 
which for $d=4$ reduces to a  topological surface term. In \req{DCYMDF} there is also the higher order correction $R^3$, which in the bosonic closed string \req{Metsaev} originates  at quadratic order in $\ap^2R^3$.  It should be
emphasized that in higher--point gravitational  amplitudes the order $\ap^2$  cannot be reproduced from the double copy of the $\ap$ order of the corresponding open bosonic string amplitudes \req{OBS} due to additional $\ap^2\zeta_2F^4$ contributions from a single  open string sector  \cite{Broedel:2012rc,Garozzo:2018uzj,Ahmadiniaz:2021ayd}. 

It is worth mentioning that  HD gravity may be related to  bimetric gravity. The latter is constructed through
interactions of two spin two fields $g_{\mu\nu}$ and $f_{\mu\nu}$ with a nonlinear interacting potential \cite{Has2011}. It has been shown \cite{CGBi} that by expanding the potential of bimetric gravity around a specific solution of $f_{\mu\nu}$
in terms of $g_{\mu\nu}$ and integrating   out $f_{\mu\nu}$ one can match the  Lagrangian of bimetric gravity to that of  HD gravity. Therefore, the intersection number associated with the amplitude in (\ref{AMPHD}) can also be related to some particular parameters of the integrated bimetric gravity known as partially massless bimetric gravity, cf. Subsection \ref{PMBI}.

\subsect{Bimetric theory as a double copy}
\label{PMBI}

The construction of the double copy \req{DCYMDF} may be related to some limit of bimetric gravity. The starting point of this map is the connection between  conformal gravity and bimetric gravity. Looking at the bimetric potential and its parameters one can find a region in the parameter space of bimetric theory known as partially massless (PM) \cite{CGBi}. In this parameter configuration, the bimetric theory in quadratic order in the curvature $R_{\mu\nu}$ matches conformal gravity. 

Bimetric gravity is a consistent and ghost--free theory that consists of two spin--two fields, i.e. two metric tensors. The action for ghost free bimetric gravity was written by Hassan and Rosen in $d=4$ dimension as \cite{Has2011}:
\begin{equation} \label{HR}
S= m_g^2 \Bigg(\int d^4 x\; \sqrt{g}\; R(g) +  \alpha^2  \int d^4 x \sqrt{f}\; R(f)  - 2\, m^2  \int d^4 x\; \sqrt{g}\; V(S;\beta_n)\Bigg)\ .
\end{equation}
Above, we have two independent Einstein--Hilbert kinetic terms for the two spin--two fields $g_{\mu\nu}$ and $f_{\mu\nu}$ with two different Planck masses $m_g$ and $m_f$ (with $\alpha=\frac{m_f}{m_g}$) and $m$ is the mass scale of the massive spin--two field after diagonalization of the two metrics. These two spin--two fields interact via  the potential $V(S;\beta_n)$. This potential couples the two metrics and creates non--trivial self and mixed interactions and is given by
\begin{equation} \label{pot}
V(S;\beta_n) =  \sum\limits_{n=0}^4 \beta_n\; e_n(S) \ , 
\end{equation}
where $\beta_n$ are five free dimensionless parameters and the symmetric polynomials $e_n(S)$ depend on the matrix $S^\mu_{\hphantom{\mu}\nu}$. Specifically, we have:
\begin{equation} 
\begin{aligned}
 e_n(S)&= S^{\mu_1}_{\hphantom{\mu_1} [\mu_1} \dots S^{\mu_n}_{\hphantom{\mu_n} \mu_n]} \ ,\\
 S^\mu_{\hphantom{\mu}\nu} S^\nu_{\hphantom{\mu}\rho}& = g^{\mu \sigma} f_{\sigma \rho} \quad \textrm{namely:}  \quad S^\mu_{\hphantom{\mu}\nu} = (\sqrt{g^{-1}f})^\mu_{\hphantom{\mu}\nu}   \ .
\end{aligned}
\end{equation}
In the action (\ref{HR}) we have two independent Ricci tensors associated with the two metrics $g_{\mu\nu}$ and $f_{\mu\nu}$, respectively. Therefore, we  have two independent diffeomorphism transformations which keep each of the two Ricci tensors invariant. There is a special limit of this theory known as partially massless (PM) bimetric theory,  which is defined over the following background solution in order to eliminate the $f_{\mu\nu}$ \cite{CGBi}
\begin{equation} 
\begin{aligned}
& f_{\mu\nu}=a^2 g_{\mu\nu}+\frac{2\gamma}{m^2} P_{\mu\nu}+ \mathcal{O}(m^{-4})\ , \label{solution}
\end{aligned}
\end{equation}
where $P_{\mu\nu}$ is the Schouten tensor. Plugging these relations back into (\ref{pot}), in \cite{CGBi} it is shown that the  kinetic term of the second metric $f_{\mu\nu}$ and the bimetric potential assume the following form (in arbitrary dimension):
\begin{equation}
\begin{aligned}
m^2 \sqrt{g}V(S;\beta_n)&= m^2 \sqrt{g}\sum\limits_{n=0}^4 \beta_n e_n(S)\sim \Lambda_0\; \sqrt{g} [1+ \kappa R(g)]\\
&+\sqrt{g} \frac{\Lambda_1}{m^2}\;\lf(\frac{d}{4(d-1)}R^2-R^{\mu\nu}R_{\mu\nu}\ri) +\mathcal{O}(R^3)\ , \\
 \sqrt{f}R(f)&\sim(\Lambda_2)^{d-2} \sqrt{g} R(g)-\frac{\Lambda_3}{m^2} \sqrt{g}\;\lf(\frac{d}{4(d-1)}R^2-R^{\mu\nu}R_{\mu\nu}\ri) +\mathcal{O}(R^3) \ .\label{twocg}
\end{aligned}
\end{equation}
 The parameters $\Lambda_i (\beta_n)$ are dimensionless. For simplicity we absorbed other dimensionless parameters, e.g. $\gamma$ in (\ref{solution}), in the parameters $\Lambda_i$, for details of the calculation we refer to \cite{CGBi}. We fix the parameters $\beta_n$ in such a way that the term including the cosmological constant vanishes i.e. $\Lambda_0=0$. For the full fledged PM bimetric one has to fix all remaining $\beta_n$ so that also $\Lambda_2(\beta_n)=0$ \cite{CGBi}.
The last two terms of \req{twocg} correspond to the CG Lagrangian (\ref{cg}). More concretely, one has the following identity, cf. \req{WeylGB}
\begin{equation}
(W_{\mu\nu\alpha\beta})^2=R_{GB}+\Big(\frac{d}{4(d-1)}R^2-R^{\mu\nu}R_{\mu\nu}\Big)\ ,
\end{equation}
 with $R_{GB}$  the Gauss-Bonnet term, which is a total derivative in four dimension and can be discarded  in the Lagrangian. After putting for $d=4$ the relations  \req{twocg} back into \req{HR} we evidence  that there are two copies of CG together with the kinetic term for the first metric $g_{\mu\nu}$ described by GR. Concretely, we obtain:
\begin{equation}
\begin{aligned}
 \mathcal{L}&= m_g^2\; \sqrt{g}\; \Bigg[ R(g) + \al^2\Lambda_2^{2} R(g)-\al^2\frac{\Lambda_3}{m^2}\;\lf(\fc{1}{3}R^2-R^{\mu\nu}R_{\mu\nu}\ri)  \\ 
 & - 2\,\frac{\Lambda_1}{m^2}\;\lf(\fc{1}{3}R^2-R^{\mu\nu}R_{\mu\nu}\ri)\ \Bigg] +\mathcal{O}(R^3)\ . \label{PMaction}
  \end{aligned}
\end{equation}
Combining the terms \req{PMaction} into the action  we reproduce \req{HDGR} with:
\begin{equation}
\begin{aligned}\label{lambdas}
 & \lambda_{1}= (1+ \al^2\Lambda_{2}^{2})\ , \\
 & \lambda_{3}=-\frac{1}{m^2}\;( 2\Lambda_1+\al^2\Lambda_3)\ .
\end{aligned}
\end{equation}
Then, the resulting Lagrangian \req{HDGR} describes a  HD gravity with the  higher order
corrections due to the bimetric theory.  Calculating the equations of motions for the action
\req{HDGR} in the leading order\footnote{Note, that $R^3$ is of order $m^4$.} of $m^2$ one ends up with the same constraints \req{eqhd} as in EW, cf. e.g. \cite{Ferrara:2018wqd} and the comment below Eq. \req{discuss2}. 

Recall, that we have eliminated $f_{\mu\nu}$ through (\ref{solution}) and obtained the Lagrangian \req{HDGR}. 
Similar to EW the PM bimetric gravity propagates   seven degrees of freedom (packaged into the metric $g_{\mu\nu}$), which contain a massless spin--two and a massive spin--two. In PM bimetric gravity the scattering amplitude ${\cal A}(n)$ describes the interaction of $n$ spin--two fields $g_{\mu\nu}$. The latter is described by the Lagrangian \req{HDGR} comprising the conformal term+GR+$R^3$ corrections.  Eventually, the amplitude ${\cal A}(n)$ may be computed by using \req{intersection} and is given by the intersection number \req{AMPHD}.
Looking at  Table \ref{t5} we may anticipate for PM bimetric gravity the double copy structure \req{DCYMDF}. 
In addition to GR in \req{DCYMDF} there are the two CG theories originating in  (\ref{PMaction}).  The latter can  be combined in one single subject to \req{lambdas}, while the $R^3$ term is of higher order $m^4$.

A final comment is, that choosing the potential  (\ref{pot}) as
\begin{equation}
V(\beta; S)\approx\ m^2\;[(f^\mu_\mu)^2-f^{\mu\nu}f_{\mu\nu}]
\end{equation}
leads to Einstein--Weyl gravity. 
This has been pointed out in \cite{Fer} and  also commented below Eq. \req{discuss2}. 
Finally, a detailed discussion on string scattering amplitudes for full bimetric gravity with comments on their double copy structure can be found in \cite{Lust:2021jps,LMMS}.

\sect{Amplitude relations in intersection theory}\label{AMPREL}

In this section we describe amplitude relations in the language of twisted forms. Furthermore, we shall look at the EYM amplitude and expand the latter in terms of an orthonormal basis of twisted forms describing  YM amplitudes. 

\subsection{Amplitude relations from  intersection theory}\label{AmpsRel}

It has been emphasized  in \cite{M1} that the fundamental KK and BCJ relations  in the language of twisted intersection theory may be understood as the total derivative (exact twisted form) associated to the color form $\varphi^{color}_{\pm,n-1}$ i.e.:
\begin{equation}
(d \pm d\omega \wedge )\,\varphi^{color}_{\pm;n-1}\sim 0 \in H^{n-3}_{\omega}\ . \label{BCJKK}
\end{equation} 
Recall, that we are using both $\varphi^{color}$ and $PT(1,2,\ldots,n)$ for the color form.
To derive amplitude relations in intersection theory  we first note that $d\omega$ corresponds to the scattering equations \req{SEQ}, i.e.:
\begin{equation}
d\omega= \sum\limits_{i=2}^{n-2} S_i\; dz_i\ , \qquad \textit{with:} \qquad S_i=\sum\limits^{n-2}_{j \neq i}\frac{s_{ij}}{z_{ij}}\ .
\end{equation} 
Secondly, to simplify  the calculation  we introduce the insertion function $\mathrm{Ins}(i)_{jk}$s:
\begin{equation}
\mathrm{Ins}(i)_{j,k}:=\frac{z_{jk}}{z_{ji}z_{ik}} \ ,
\ee
which acts on the color form $PT(1,\ldots,n)$ as operator:
\be
\mathrm{Ins}(i)_{j,k}\; PT(1,2,3,\ldots,j,k,...,n)=PT(1,2,3,\ldots,j,i,k,...,n)\ .
\end{equation} 
Therefore, by using momentum conservation  and performing some rearrangements    we can write  $S_i$ in terms of $\mathrm{Ins}(i)_{j,k}$ as
\begin{equation}
 S_i=\sum\limits^{n-2}_{j \neq i}x(i)_j\;\mathrm{Ins}(i)_{j,j+1} \ ,
 \end{equation} 
where $x(i)_j$ is defined as:  
\begin{equation}
x(i)_j=p_i \cdot \sum_{k=2}^j p_k\ .
\end{equation}
Now, by using  $\varphi^{color}_{\pm;n-1}=PT(1,2,\ldots,n-1)$ and  $d\varphi^{color}_{\pm;n-1}=0$ we rewrite  (\ref{BCJKK}) in the following way:
\begin{equation}
  \begin{aligned}
\Phi&:=(d \pm d\omega \wedge )\;\varphi^{color}_{\pm;n-1}=\pm(d\omega \wedge )\;\varphi^{color}_{\pm;n-1}\\[2mm]
&=(d\omega\big|_{dz_n} \wedge )\varphi^{color}_{\pm;n-1}=(S_n dz_n\wedge )\varphi^{color}_{\pm;n-1}\simeq 0 \, \in H^{n-3}_{\omega} \ .
  \end{aligned}\label{MizBCJ}
\end{equation} 
Above $d\omega|_{dz_n}$ projects onto the  one--form part $dz_n$ of $d\omega$.
Eventually, from \req{MizBCJ} we deduce the following relation
\be
\sum\limits^{n-2}_{j \neq n}x(n)_j\;\mathrm{Ins}(n)_{j,j+1}PT(1,2,3,...,n-1)=\sum\limits^{n-2}_{j \neq n}x(n)_j\; PT(1,2,3,..,j,n,j+1...,n-1)=0,
\ee
which assumes the form of a BCJ relation.
To produce the BCJ relations for YM amplitudes we shall calculate the fowling intersection number
\begin{equation}
  \begin{aligned}
&\langle \varphi^{gauge}_{-,n},\Phi\rangle_{\omega}=\langle \varphi^{gauge}_{-,n},(S_n dz_n\wedge )\varphi^{color}_{\pm;n-1}\rangle_{\omega}\\
&=\langle \varphi^{gauge}_{-,n}, \sum\limits^{n-2}_{j \neq n}x(n)_j PT(1,2,3,..,j,n,j+1...,n-1)\rangle_{\omega}\\
&= \sum\limits^{n-2}_{j \neq n}x(n)_j\; \underbrace{\langle \varphi^{gauge}_{-,n},  PT(1,2,3,..,j,n,j+1...,n-1)\rangle_{\omega}}_{ \mathcal{A}_{YM}(1,2,3,..,j,n,j+1...,n-1)}=0 \ ,
  \end{aligned}
\end{equation} 
from which the BCJ relation for YM amplitudes follows: 
\be
\sum\limits^{n-2}_{j \neq n}x(n)_j\;   \mathcal{A}_{YM} (1,2,3,..,j,n,j+1...,n-1)=0\ .
\end{equation} 

On the other hand, for the KK relations we start at the $n+1$--form
 $$\varphi^{color}_{\pm,n+1}(\sigma_l):=PT(1,2,...,l,p,l+1,....,n)\ ,$$ 
 with $n+1$ legs. One additional leg denoted by $p\equiv n+1$ is appended to $n$ legs such that there are in total $n+1$ legs. Furthermore  $\sigma_l$ denotes the particular ordering of those $n+1$ legs as: $\sigma_l\leftrightarrow(1,2,...,l,p,l+1,....,n)$.
As a consequence of KK relations for PT factors we have the identity:
\begin{equation}\label{MizKK}
  \begin{aligned}
\sum\limits_{l=1}^{n} \varphi^{color}_{\pm,n+1}(\sigma_l) = \sum\limits_{l=1}^{n}  PT(1,2,\ldots,l,p,l+1,....,n)=0\ .
  \end{aligned}
\end{equation}
Inserting  \req{MizKK} into the intersection number \req{intersection} appended by 
$\varphi^{gauge}_{-,n+1}$ yields
\begin{equation}
 \langle \varphi^{gauge}_{-,n+1},\sum\limits_{l=1}^{n} \varphi^{color}_{+,n+1}(\sigma_l)\rangle_{\omega}=  \sum\limits_{l=1}^{n} \langle \varphi^{gauge}_{-,n+1},\varphi^{color}_{+,n+1}(\sigma_l)\rangle_{\omega}=0\ ,
 \ee
from which the KK relation for YM amplitudes of generic helicity configurations follows: 
 \be 
 \sum\limits_{l=1}^{n} \mathcal{A}_{YM}(1,2,\ldots,l,p,l+1,....,n)=0\ .
\end{equation}

\subsect{Expansion Coefficient for EYM amplitude}

Any intersection number \req{intersection} may be expanded w.r.t. a orthonormal basis of $n$--forms $\bigcup\limits_{a=1}^{(n-3)!}\{\Phi_{+,a}\}\in H_{+\om}^{n-3}$. 
Generally, together with the dual basis $\bigcup\limits_{b=1}^{(n-3)!}\{\Phi^\vee_{-,b}\}\in H_{-\om}^{n-3}$ with the intersection matrix
\be\label{onb}
\vev{\Phi_{+,a},\Phi^\vee_{-,b}}_\om=\delta_{ab}\ ,
\ee
we have the following expansion of twisted $n$--forms   $\varphi^1_{+},\varphi^2_{-}$
\begin{align}
 \varphi^1_{+}&=\sum_{a=1}^{(n-3)!}\vev{\Phi^\vee_{-,a},\varphi_{+} }_\om \ \Phi_{+,a}\ ,\\
 \varphi^2_{-}&=\sum_{b=1}^{(n-3)!}\vev{\varphi_{-}   , \Phi_{+,b}}_\omega\ \Phi^\vee_{-,b}\ ,
\end{align}
respectively, cf. \req{EXPF} for the case of $\Phi_{+,a}=PT(a)$ leading to \req{DCE}:
\be\label{ODE}
\vev{\varphi_+^1,\varphi^2_-}=\sum_{a=1}^{(m-3)!}\vev{PT^\vee(a),\varphi_+^1}_\om\ \vev{\varphi_-^2,PT(a)}_\om\ .
\ee
With $\varphi^1_+=\tilde\varphi^{EYM}_{+,n;r}, \varphi^2_-=\varphi^{EYM}_{-,n;r}=\varphi^{gauge}_{-,n+r}$ the  orthogonal decomposition \req{ODE} can be used to express EYM amplitudes in terms of a linear combination of a basis of $m:=n+r$--point YM subamplitudes \cite{MS}: 
\be\label{resultEYM}
\mathcal{A}_{EYM}(n;r)=\lim_{\ap \rightarrow \infty} \
  \sum_{a=1}^{(m-3)!}
    \vev{PT^\vee(a), \widetilde{\varphi}^{EYM}_{+,n;r} }_\om \  \ \Ac_{YM}(a)\ .
 \ee
In addition, we may use  BCJ--KK relations for further simplifications. 

To determine in \req{resultEYM} the coefficients $\vev{PT^\vee(a), \widetilde{\varphi}^{EYM}_{+,n;r} }_\om$ 
we exemplify the one graviton case $r=1$ and we label the momentum of this one graviton by $p\equiv n+1$. From  \cite{MS} we have for the EYM twisted form $\tilde{\varphi}^{EYM}_{n;1}$
\begin{equation}
\begin{aligned}
\lim\limits_{\alpha' \rightarrow \infty} \tilde{\varphi}^{EYM}_{n;1}&=d\mu_{n+1}\;PT(1,\ldots,n) \;\sum\limits_{l=1}^{n-1} (\varepsilon_{p} \cdot x_l) \frac{z_{l,l+1}}{z_{l,p}z_{p,l+1}}\\
&=d\mu_{n+1}\;\sum\limits_{l=1}^{n-1} (\varepsilon_{p} \cdot x_l)\;PT(1,2,...,l,p,l+1,...,n)\ ,\label{exprEYM}
\end{aligned}
\end{equation}
with the definition  $x^\mu_l=\sum\limits_{j=1}^{l} k^\mu_j$.
Inserting \req{exprEYM} into (\ref{resultEYM}) yields:
\begin{equation}
\begin{aligned}
 \mathcal{A}_{EYM}(n;1) &=\lim\limits_{\alpha' \rightarrow \infty}\sum\limits_{\al\in S_{n-3}} \langle PT^{\vee}(\al), \tilde{\varphi}^{EYM}_{n;1} \rangle\; \mathcal{A}_{YM}(\al)\\
& = \sum\limits_{\al\in S_{n-3}}\sum\limits_{l=1}^{n-1} (\varepsilon_{p} \cdot x_l)\; \langle PT^{\vee}(\al), PT(1,2,...,l,p,l+1,...,n) \rangle_{\omega}\; \mathcal{A}_{YM}(\al)\ .\label{zwischen}
\end{aligned}
\end{equation}
Notice, that above the $\ap$--dependence has dropped. Now we can use the orthonormality condition \req{onb}, i.e. 
\begin{equation}
\langle PT^{\vee}(\alpha), PT(\beta) \rangle_{\omega}=\delta_{\alpha,\beta}\ , \label{oriti} 
\end{equation}
label  the ordering of $PT(1,2,...,l,p,l+1,...,n)$ by $\sigma_l$ and \req{zwischen} becomes
\begin{equation}
\begin{aligned}
\mathcal{A}_{EYM}(n;1)& = \sum\limits_{\al\in S_{n-3}}\sum\limits_{l=1}^{n-1} (\varepsilon_{p} \cdot x_l)\; \langle PT^{\vee}(\al), PT(\sigma_l) \rangle_{\omega}\; \mathcal{A}_{YM}(\al)\ .\\
& = \sum\limits_{\al\in S_{n-3}}\sum\limits_{l=1}^{n-1} (\varepsilon_{p} \cdot x_l)\; \delta_{\alpha,\sigma_l}\; \mathcal{A}_{YM}(\al)\\
& =\sum\limits_{l=1}^{n-1} (\varepsilon_{p} \cdot x_l)\; \mathcal{A}_{YM}(1,2,...,l,p,l+1,...,n)\ ,
\end{aligned}
\end{equation}
which is the relation from \cite{ST}.

Next, let us also evaluate  the coefficients $\vev{PT^\vee(a), \widetilde{\varphi}^{EYM}_{+,n;r} }_\om$ in \req{resultEYM} for the two graviton case $r=2$.
Again, the starting point is the  twisted form  $\tilde{\varphi}^{EYM}_{n;2}$ for EYM
given  in \cite{MS}
\begin{align}
\lim_{\alpha' \rightarrow \infty}\widetilde{\varphi}^{EYM}_{\pm,n;2}&= d\mu_{n+2}\,\,PT(1,2,3,...,n)\int  \prod\limits_{ i\in\Sc_2}   d \theta_i d\bar{\theta_i} \ \exp\Bigg\{\frac{1}{2} \ap^2\sum\limits_{i,j\in \Sc_2} \begin{pmatrix} \theta_i \\ \bar{\theta}_i  \end{pmatrix}^t \Psi_{\Sc_2}  \,\, \begin{pmatrix} \theta_j \\ \bar{\theta}_j  \end{pmatrix}\Bigg\}\Bigg|_{\ov{\zeta_l} \rightarrow \zeta_l}\nonumber\\
& =d\mu_{n+2}\,\, PT(1,2,3,...,n)\; \pf' \Psi_2\ ,\label{Zurich}
\end{align}
with $\Sc_2=\{n+2,n+4\}$.
To compute the coefficient $\langle PT_a^{\vee}, \tilde{\varphi}^{EYM}_{n;2} \rangle$ of the decomposition  \req{resultEYM}  we expand the twisted form \req{Zurich} in terms of a PT basis (for a derivation we refer to \ref{A3}) we have: 
\begin{equation}
\begin{aligned}
 \lim\limits_{\alpha' \rightarrow \infty} \tilde{\varphi}^{EYM}_{n;2}=d\mu_{n+2} & \,\Bigg\{\sum\limits_{k \geq l=1\atop }^{n-1} (\varepsilon_{q} \cdot x_l) (\varepsilon_{p} \cdot x_k)PT(1,2,...,l,p,l+1,...k,q,k+1,...,n)+(p \leftrightarrow q) \\
&\quad - \sum\limits_{l=1}^{n-1} (\varepsilon_{p} \cdot q)(\varepsilon_{q} \cdot x_l) \sum\limits_{j=2}^{l-1} PT(1,j-1,q,j,...l-1,p,l,...,n)+ (p \leftrightarrow q)\\
&\qquad -\frac{1}{2}\sum_{l=1}^{n-1} (\varepsilon_{p} \cdot \varepsilon_{q} ) s_{p,l}\sum\limits_{j=2}^{l} PT(1,j-1,q,p,j,...,l,...,n)+(p \leftrightarrow q)\Bigg\}\ . \label{expresult}
\end{aligned}
\end{equation}
Now we proceed similar to the one graviton case and insert \req{expresult} into \req{resultEYM} and apply the orthonomality condition (\ref{oriti}) to get:
\begin{align}
 \Ac_{EYM}(n;2) =&\lim\limits_{\alpha' \rightarrow \infty}\sum\limits_{a=1}^{(n-3)!} \langle PT^{\vee}(a), \tilde{\varphi}^{EYM}_{n;2} \rangle\; \Ac_{YM}(a) \label{resPle} \\
 =& \sum\limits_{k \geq l=1\atop }^{n-1} (\varepsilon_{q} \cdot x_l) (\varepsilon_{p} \cdot x_k)\Ac_{YM} (1,2,...,l,p,l+1,...k,q,k+1,...,n)+(p \leftrightarrow q)  \nn\\
& -\sum\limits_{l=1}^{n-1} (\varepsilon_{p} \cdot q)(\varepsilon_{q} \cdot x_l) \sum\limits_{j=2}^{l-1} \Ac_{YM} (1,j-1,q,j,...l-1,p,l,...,n)+ (p \leftrightarrow q)\nn\\
& \qquad-\frac{1}{2}\sum_{l=1}^{n-1} (\varepsilon_{p} \cdot \varepsilon_{q} ) s_{p,l}\sum\limits_{j=2}^{l} \Ac_{YM} (1,j-1,q,p,j,...,l,...,n)+(p \leftrightarrow q)\ . \nn
\end{align}  
This result provides the expansion of the EYM amplitude involving $n$ gluons and two gravitons in terms of pure YM $n+2$--point subamplitudes. The expression  \req{resPle} has been already given in \cite{p}, while here we had applied twisted intersection theory to derive it.

\sect{Concluding remarks}

While in \cite{MS} we have focused on EYM amplitudes and their description in terms of twisted intersection theory, in this work  we have extended and generalized these results into several directions. By pairing various differential forms we have computed intersection numbers \req{intersection} describing the amplitudes of various field--theories, cf. Tables \ref{t1}--\ref{t4}.
First, we have reviewed known constructions of field--theory amplitudes through their pairs of twisted forms entering the intersection number \req{intersection}, cf. Table \ref{t1}. In particular, theories 
which are described by the twisted form of EYM \req{TWF1} are tabulated in Tables \ref{t1b} and \ref{t2}.
Secondly, for known field--theories (e.g. EM, DBI and YMS) yet without any twisted from description their pairs of twisted forms has been given. In fact, one of the twisted forms of these theories is the twisted form \req{EM} for EM, cf. Table \ref{t4}. On the other hand, a new twisted form \req{phiB} has been extracted from the twisted form $\varphi^{bosonic}$ related to the bosonic string. This form is relevant 
for describing amplitudes \req{CHY DFPh} of $(DF)^2$ theory.
Furthermore, we have applied the embedding from the disk onto the sphere to  bosonic open string fields to construct a novel twisted form \req{embW}. The latter allows to construct amplitudes by intersection numbers for a variety of theories including Weyl--YM or $(DF)^2+\phi^3$ theories, cf. Table \ref{tt4}.
In Section~\ref{NDC} we have elaborated  on the generic underlying structure of double copy construction in intersection theory. In particular, we have been concerned with theories whose pairs of  twisted differentials entering \req{intersection} contain a color form $\varphi^{color}$. For those theories CK duality is easily exhibited and hence they serve as natural candidates for double copies, cf. also Table \ref{tnew}.
Building up on this construction in  Tables \ref{t5} and \ref{t6} we have  presented old and new double copies. For the latter  an amplitude description has been unknown previously. In particular, in Table \ref{t7} we present a collection of previously unknown spin--two double copy constructions. 
We find a double copy construction for HD gravity, (partial massless) bimetric gravity and some more exotic theories. Our findings are not complete and it would be very interesting to build up on those and extending the list of possible new double copies. In Section 6 we present a derivation of Kleiss--Kuijf (KK) and Bern--Carrasco--Johansson  amplitude relations in intersection theory. 
Finally, the appendices contain some supplementary material.

While throughout this work we have only discussed tree--level amplitudes and their underlying twisted intersection description a natural further step is to extend our results to higher--loops.
A subsequent question is finding appropriate  twisted forms on the elliptic curve which generalize \req{listT} and give rise to one--loop integrands of gauge and gravity theories. The one--loop monodromy results  from \cite{Hohenegger:2017kqy,Casali:2019ihm,Stieberger:2021daa} are of importance  for constructing such differential forms.
The KLT relations  \cite{klt} are the foundation of any double copy relation at tree--level
both in the CHY formalism and twisted intersection theory.
Consequently, one--loop KLT relations  on the Riemann torus \cite{Stieberger:2022lss} should trigger
one--loop double copies in twisted intersection theory.

\section*{Acknowledgment}
One of the authors (PM) would like to thank Chrysoula Markou for useful discussions and  suggestions  and Hrólfur Ásmundsson for very useful discussions. 
%\newpage

\appendix

\section{EYM basis expansion}
\label{A3}

Here we  expand the twisted from $\hat{\tilde{\varphi}}^{EYM}_{n;2}$
\begin{equation}
    \begin{aligned}
\lim_{\alpha' \rightarrow \infty}\widetilde{\varphi}^{EYM}_{\pm,n;2}=d\mu_{n+2}\,\, PT(1,2,3,...,n)\; \pf' \Psi_2\ 
   \end{aligned}
\end{equation}
w.r.t. a Park--Taylor basis. 
We start with the matrix $\Psi_{\Sc_2}$:
 \begin{equation}
    \begin{aligned}
 & \Psi_{\Sc_2}=\begin{pmatrix} 
    0 & \frac{p \cdot q}{\ov{z}_{3,4}} & -\Psi^{4}_{1} & \frac{\widetilde{\varepsilon}_{q} \cdot p}{\ov{z}_{3,4}} 
 \\   -\frac{p \cdot q}{\ov{z}_{3,4}} & 0 & \frac{-\widetilde{\varepsilon}_{p} \cdot q}{\ov{z}_{3,4}}  & -\Psi^{6}_{1}
 \\   \Psi^{4}_{1} &  \frac{\widetilde{\varepsilon}_{p} \cdot q}{\ov{z}_{3,4}}  & 0 &  \frac{\widetilde{\varepsilon}_{p} \cdot \widetilde{\varepsilon}_{q}}{\ov{z}_{3,4}} 
 \\   -\frac{\widetilde{\varepsilon}_{q} \cdot p}{\ov{z}_{3,4}} & \Psi^{6}_{1} & -\frac{\widetilde{\varepsilon}_{p} \cdot \widetilde{\varepsilon}_{q}}{\ov{z}_{3,4}}  & 0
 \end{pmatrix}\ .
    \end{aligned}
\end{equation}
The Pfaffian of this matrix is given by
\begin{equation}
\begin{aligned}
\pf \Psi_{\Sc_2}& = \Psi^{p}_{1}\Psi^{q}_{1}-\frac{s_{pq} (\varepsilon_p \cdot \varepsilon_q) }{z_{pq}^2}+\frac{(\varepsilon_p \cdot q) (\varepsilon_q \cdot p)}{z_{pq}^2}\ ,
\end{aligned} \label{expands2}
\end{equation}
with the following definitions:
\begin{equation}
\begin{aligned}
&  \Psi^p_{1}=\sum\limits_{l=1}^{n-1} (\varepsilon_p \cdot  x_l)\ \frac{\sigma_{l,l+1}}{\sigma_{l,p}\sigma_{l+1,p}} + (\varepsilon_p \cdot q)\ \frac{\sigma_{q,n}}{\sigma_{p,q}\sigma_{n,p}}= \tilde{\Psi}^p_{1}  + (\varepsilon_p \cdot q)\ \frac{\sigma_{q,n}}{\sigma_{p,q}\sigma_{n,p}}\ , \\
&  \Psi^q_{1}=\sum\limits_{l=1}^{n-1} (\varepsilon_q \cdot  x_l)\ \frac{\sigma_{l,l+1}}{\sigma_{l,q}\sigma_{l+1,q}} + (\varepsilon_q \cdot p)\ \frac{\sigma_{p,n}}{\sigma_{q,p}\sigma_{n,q}}=\tilde{\Psi}^q_{1} + (\varepsilon_q \cdot p)\ \frac{\sigma_{p,n}}{\sigma_{q,p}\sigma_{n,q}}\ , \\
& x_l=\sum\limits_{m=1}^l k^\mu_m \ ,\\
\end{aligned} \label{cqq}
\end{equation}
where we denote the first summands in the above terms by $\tilde{\Psi}^p_1$ and $\tilde{\Psi}^q_1$, respectively.
Using these and plugging them back into (\ref{expands2}) we can cancel the the last term in (\ref{expands2}), which is a double pole, and we obtain:
\begin{equation}
\begin{aligned}
\pf \Psi_{\Sc_2}& = \tilde{\Psi}^{p}_{1}\tilde{\Psi}^{q}_{1}+ \tilde{\Psi}^{p}_{1} (\varepsilon_q \cdot p) \frac{z_{pn}}{z_{pq}z_{qn}}+ \tilde{\Psi}^{q}_{1} (\varepsilon_p \cdot q) \frac{z_{qn}}{z_{pq}z_{pn}} -\frac{s_{pq}\varepsilon_p \cdot \varepsilon_q}{z_{pq}^2}\ .
\end{aligned}
\end{equation}
Now we can write the twisted from $\hat{\tilde{\varphi}}^{EYM}_{n;2}$ in the following way:
\begin{equation}
\begin{aligned}
& \lim\limits_{\alpha' \rightarrow \infty} \hat{\tilde{\varphi}}^{EYM}_{n;2}=PT(1,2,...,n)\pf \Psi_{\Sc_2} \\
& =PT(1,2,...,n)\Bigg[ \tilde{\Psi}^{p}_{1}\tilde{\Psi}^{q}_{1}+ \tilde{\Psi}^{p}_{1} (\varepsilon_q \cdot p) \frac{z_{pn}}{z_{pq}z_{qn}}+ \tilde{\Psi}^{q}_{1} (\varepsilon_p \cdot q) \frac{z_{qn}}{z_{pq}z_{pn}} -\frac{s_{pq}\varepsilon_p \cdot \varepsilon_q}{z_{pq}^2}\Bigg]\ .
\end{aligned} 
\end{equation}
At this point we are able to expand this expression in terms of a Park--Taylor basis and use the orthogonality  condition of that basis to obtain amplitude relations. In order to expand we  use the following relations (cf. for the proof \cite{p}): 
\begin{equation}
\begin{aligned}
\tilde{\Psi}^p_{1} \tilde{\Psi}^q_{1}&=  \sum\limits_{l=1  }^{n-1} \sum\limits_{k=1 }^{n-1}(\varepsilon_p \cdot  x_l)\, (\varepsilon_q \cdot x_k) \frac{\sigma_{l,l+1}}{\sigma_{l,p}\sigma_{l+1,p}}\  \ \frac{\sigma_{k,k+1}}{\sigma_{k,q}\sigma_{k+1,q}}\ ,\\
\tilde{\Psi}_1^p (\varepsilon_q \cdot p) \frac{z_{pn}}{z_{pq}z_{qn}} PT(1,2,3,...,n)&=-(\varepsilon_q \cdot p) \sum\limits_{l=1}^{n-1} \varepsilon_p \cdot x_l \sum\limits_{\sigma=q \cup \{1,2,...,l\}} PT(\sigma,p,l+1,...,n)\ , \\
\frac{s_{pq}}{z_{p,q}^2}&=\sum\limits_{i=l}^{n-1} s_{pl} \frac{1}{z_{lp}z_{pq}} \ .
\end{aligned}  \label{relations}
\end{equation}
The last two relations are proven through implantation of multiple BCJ-KK relations and scattering equations. 
Using the first relation in (\ref{relations}) we derive:
\begin{equation}
\begin{aligned}
 \lim\limits_{\alpha' \rightarrow \infty} \hat{\tilde{\varphi}}^{EYM}_{n;2}=
 &  \sum\limits_{k \geq l=1\atop }^{n-1} (\varepsilon_{q} \cdot x_l) (\varepsilon_{p} \cdot x_k)PT(1,2,...,l,p,l+1,...k,q,k+1,...,n)+(p \leftrightarrow q) \\
& + \tilde{\Psi}_1^p (\varepsilon_q \cdot p) \frac{z_{pn}}{z_{pq}z_{qn}} PT(1,2,3,...,n)+ (p \leftrightarrow q)\\
& -\frac{(\varepsilon_{p} \cdot \varepsilon_{q} ) s_{p,q}}{z_{p,q}^2}\  
PT(1,2,3,...,n)\ .
\end{aligned} \label{expansion EYM}
\end{equation}
we can simplify this equations further
using the second and third relations in (\ref{relations}) after multiply terms with the Park-Taylor we obtain:
\begin{equation}
\begin{aligned}
 \lim\limits_{\alpha' \rightarrow \infty} \tilde{\varphi}^{EYM}_{n;2}=d\mu_{n+2} & \,\Bigg\{\sum\limits_{k \geq l=1\atop }^{n-1} (\varepsilon_{q} \cdot x_l) (\varepsilon_{p} \cdot x_k)PT(1,2,...,l,p,l+1,...k,q,k+1,...,n)+(p \leftrightarrow q)  \\
& - \sum\limits_{l=1}^{n-1} (\varepsilon_{p} \cdot q)(\varepsilon_{q} \cdot x_l) \sum\limits_{j=2}^{l-1} PT(1,j-1,q,j,...l-1,p,l,...,n)+ (p \leftrightarrow q)\\
& -\frac{1}{2}\sum_{l=1}^{n-1} (\varepsilon_{p} \cdot \varepsilon_{q} ) s_{p,l}\sum\limits_{j=2}^{l} PT(1,j-1,q,p,j,...,l,...,n)+(p \leftrightarrow q)\Bigg\}
\end{aligned}
\end{equation}
This is the expansion which we use in (\ref{expresult}).

\newpage
\addcontentsline{toc}{section}{References}

\bibliography{bibcf}
\end{document}